\documentclass[10pt]{iopart}

\usepackage{iopams}
\usepackage{graphicx}
\usepackage{epstopdf}
\usepackage{cite}

\newcommand{\ket}[1]{\left\vert{#1}\right\rangle}
\newcommand{\bra}[1]{\left\langle{#1}\right\vert}

\begin{document}

\title{Creation of bipartite steering correlations by a fast damped auxiliary mode}

\author{Li-hui Sun, Kai-Kai Zhang, Feng Xia, Bin Hu, Fang Chen, Chun-Chao Yu}

\address{Institute of Quantum Optics and Information Photonics, School of Physics and Optoelectronic Engineering, Yangtze University, Jingzhou 434023, China}
\ead{sunlihui01@hotmail.com}
\vspace{10pt}
\begin{indented}
\item[]\today
\end{indented}

\begin{abstract}
We consider a three-mode system and show how steering correlations can be created between two modes of the system using the fast dissipation of the third mode.
These correlations result in a directional form of entanglement, called quantum or EPR steering. We illustrate this on examples of the interactions among damped radiation modes in an optomechanical three-mode system. By assuming that one of the modes undergoes fast dissipation, we show that the coupling of that mode to one or two other modes of the system may result in one- or two-way quantum steering. Explicit analytical results are given for the steering parameters.
We find that two modes coupled by the parametric-type interaction and damped with the same rates can be entangled but cannot exhibit quantum steering. When, in addition, one of the modes is coupled to a fast damped mode, steering correlations are created and the modes then exhibit one-way steering. The creation of the steering correlations is interpreted in the context of the variances of the quadrature components of the modes that the steering correlations result from an asymmetry in the variances of the quadrature components of the modes induced by the auxiliary mode. It is found that the fluctuations act directionally that quantum steering may occur only when the variance of the steering mode is larger that the variance of the steered mode. The scheme is shown to be quite robust against the thermal excitation of the modes if the fluctuations of the steering mode are larger than the fluctuations of the steered mode.
\end{abstract}

\vspace{2pc}

\submitto{\JPB}

\maketitle
%

\section{Introduction}

The Einstein-Podolsky-Rosen (EPR) steering or shortly quantum steering refers to a situation in which one can distinguish the role of each subsystem in the creation of entanglement~\cite{es35,wj07,jw07,cj09,rd09,rt15,gsa14}. Such a situation does not exist in entanglement which is shared equally between the subsystems, that one cannot judge which of the subsystems is responsible for the entanglement. It has been suggested that the asymmetric property of quantum steering can be used to achieve secure quantum communication~\cite{bc}, secure quantum teleportation~\cite{md13}, and quantum key distribution~\cite{ww14}.

Various systems have been theoretically studied for the creation of quantum steering~\cite{sd13,olsen13,sn14,bv14,tr14,wgh14,he15,ade14,ade15,rh16,zc17,wl17,kl17,kl17a,ct18,wx18}. The experimental realization of quantum steering has been accomplished in several systems~\cite{sj10,sg12,be12,vh12,bw12,ss13,Guo14,nc15,seiji15,ww16}. Of particular interest is the possibility to achieve quantum steering in macroscopic systems. For this reason, considerable attention has been devoted to  optomechanical systems, where the dynamics of the macroscopic mechanical mode of the oscillating mirror can be controlled through the coupling to an optical cavity mode~\cite{wc14,tz15,yl15,xs15,ts15,qd17,lz17}.

A key question for quantum steering is the effect of dissipation which in long interaction times tends to destroy the steering, particularly in macroscopic systems such as optomechanical cavities. The main sources of dissipation are the decay of the cavity photons and damped oscillations of the mirror which prevent to achieve a strong entanglement~\cite{vt07,gv07,vg07,bg08,gm08,zh11,sl12,wc13,sx17}. It has been proposed the use of pulsed excitation resulting in a short time interaction between the modes which can significantly reduce the effect of dissipation~\cite{hw11,hr13,hf14,wg14,kh14,wg14a,sm17}.

In this paper we consider a dissipative three-mode system, which decays to a mixed Gaussian state. We illustrate the three-mode system on example of a dissipative three-mode optomechanical system composed of an ensemble of two-level atoms located inside a single mode cavity with a movable mirror and driven by a coherent laser field. We show how to achieve a steady-state bipartite quantum steering via fast dissipation of one of the modes. To quantify steering, we adopt the Reid criterion~\cite{md89}, which is applicable to non-idealized Gaussian continuous variable systems, i.e. for dissipative Gaussian systems which decay to decohered mixed Gaussian states~\cite{rt15,tz15,sl12,hr13,hf14,wg14,kh14,wg14a,sm17}.
We adiabatically eliminate the fast damped mode which enables us to obtain modified equations of motion for the operators of the remaining two modes. The equations are then solved for the steady-state and the solutions are used to derive explicit analytical expressions for the variances of the quadrature components and correlations between the modes. The method of the adiabatic elimination of a fast damped mode from the dynamics of a system has been carried out in a number of previous work. For example, Parkins {\it et al.}~\cite{ps06} have used the method for the unconditional preparation of EPR-type entangled states of collective atomic modes in physically separated atomic ensembles. The method was also used to demonstrate the cavity loss induced generation of entangled atoms~\cite{ph99}.
It has also been used in optomechanical systems to demonstrate entanglement and quantum steering in pulsed excitation of a fast damped cavity mode~\cite{hw11,hr13,hf14,wg14,kh14,wg14a,sm17}.

The paper is organised as follows.
We introduce our physical system in Sec.~\ref{Sec2}, using a three-component optomechanical system composed of an ensemble of $N$ two-level atoms located inside a single mode cavity with a movable mirror and driven by a coherent laser field. In Sec.~\ref{Sec3}, we introduce parameters that measure the degree of bipartite steering, which are simply related to the variances of the quadrature components of the modes and correlations between them. Next, in Sec.~\ref{Sec4} we examine under what kind of conditions a bipartite steering can be achieved between two modes coupled to the fast damped mode, and in Sec.~\ref{Sec5} we examine the effect of the auxiliary mode on the creation of bipartite entanglement. We also discuss the effect of thermal excitation of the modes. Finally, in Sec.~\ref{Sec6} we summarize our results.

\section{The physical system}\label{Sec2}

\begin{figure}[h]
\includegraphics[width=0.5\columnwidth]{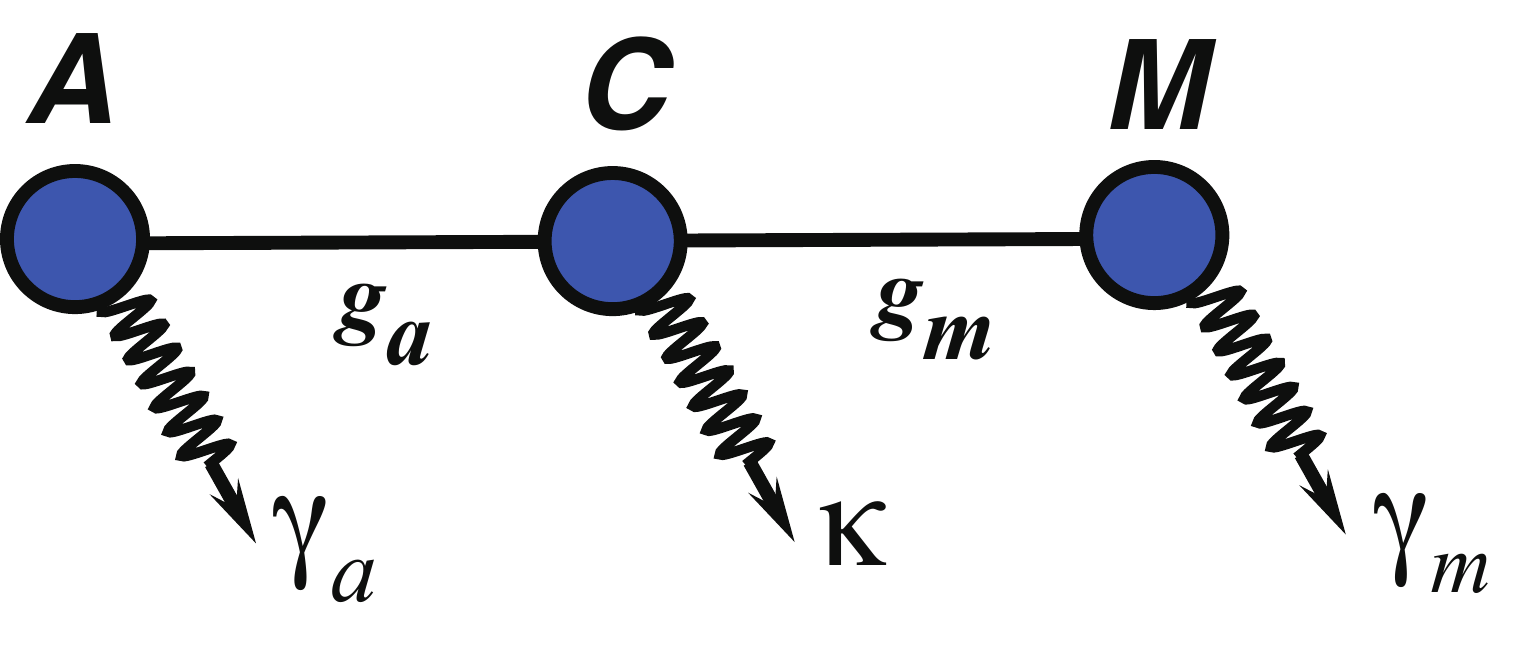}
\caption{(Color online) Schematic illustration of three harmonic oscillators, depicted as $A$, $C$, and $M$, damped with rates $\gamma_{a}$, $\kappa$, and $\gamma_{m}$, respectively. $g_{a}$ is the coupling strength between $A$ and $C$, and $g_{m}$ is the coupling strength between $C$ and $M$.}
\label{fig1}
\end{figure}

The system we discuss consists of three damped quantum harmonic oscillators, $A$, $C$, and $M$, as shown schematically in Figure~\ref{fig1}. The oscillators are characterized by frequencies, $\omega_{a}, \omega_{c}$ and $\omega_{m}$, damping rates $\gamma_{a}, \kappa$ and $\gamma_{m}$, and are represented by boson operators $c$, $a$, and $b$, respectively.  Suppose that there is a linear type coupling between the oscillators $A-C$ with the strength $g_{a}$, and a nonlinear parametric-type coupling between $C-M$ oscillators with the strength $g_{0}$. Such coupling configuration can be realized in a three-mode optomechanical system~\cite{gb06,gh09,rd11,c11,td11,rn10}. Therefore, in what follows we focus on a damped  optomechanical system composed of a single mode cavity with an oscillating mirror and an ensemble of $N$ two-level atoms located inside the cavity. The cavity mode is driven by a coherent laser field through the fixed mirror. The $j$th atom of the atomic ensemble is represented by its ground $\ket{g_{j}}$, an excited state $\ket{e_{j}}$, and the transition frequency $\omega_{a}$.

Representing the cavity mode in terms of the annihilation (creation) operators $a (a^{\dagger})$, the oscillating mirror in terms of the annihilation (creation) operators $b (b^{\dagger})$, and the atomic ensemble in terms of the collective atomic spin operators
\begin{eqnarray}
S^{-} &=& \sum_{j=1}^{N}\ket{g_{j}}\bra{e_{j}} ,\ S_{z} =\frac{1}{2}\sum_{j=1}^{N}\left(\ket{e_{j}}\bra{e_{j}} -\ket{g_{j}}\bra{g_{j}}\right) ,
\end{eqnarray}
the total Hamiltonian of the system can be written as $(\hbar\equiv 1)$:
\begin{eqnarray}
H &=& \, \Delta_{c} a^{\dagger}a +\Delta_{a}S_{z}+\omega_{m}b^{\dagger}b+g_{e}\left(S^{+}a +a^{\dagger}S^{-}\right)  \nonumber \\
&+& g_{0}a^{\dagger}a \left(b^{\dagger}+b\right) +i\left(E_{L}a^{\dagger}-E^{\ast}_{L}a\right) ,\label{l1a}
\end{eqnarray}
where $g_{e}$ is the coupling strength of the atomic ensemble to the cavity mode, $\Delta_{c}=\omega_{c}-\omega_{L}$ and $\Delta_{a}=\omega_{a}-\omega_{L}$ are the detunings, respectively, of the cavity mode frequency $\omega_{c}$ and the atomic frequency $\omega_{a}$ from the the driving field frequency $\omega_{L}$, and $E_{L}$ is the amplitude of the laser field driving the cavity mode. In writing the Hamiltonian (\ref{l1a}) we have assumed a uniform coupling strength of the atoms to the cavity mode.

Since our main concern here is the dynamics of three coupled quantum harmonic oscillators, we perform a bosonisation of the collective atomic spins operators by
means of the Holstein-Primakoff representation~\cite{hp40}, in which the collective atomic operators are described in terms of bosonic operators as
\begin{eqnarray}
S^{-} &=& \sqrt{\left(N-c^{\dagger}c\right)}\, c ,\quad S^{+} = c^{\dagger}\sqrt{\left(N-c^{\dagger}c\right)} ,\quad S_{z} = c^{\dagger}c -\frac{N}{2} ,\label{ssz}
\end{eqnarray}
where $c$ and $c^{\dagger}$ are bosonic annihilation and creation operators of the atomic mode, satisfying the usual canonical bosonic commutation relation $[c,c^{\dagger}]=1$.
Assuming that the atomic ensemble is weakly coupled to the cavity mode that the number of excited atoms in the ensemble is much smaller than the total number of atoms, i.e., $\langle c^{\dagger}c\rangle \ll N$, the operators (\ref{ssz}) can be approximated by
\begin{eqnarray}
S^{-} \approx \sqrt{N}c ,\quad  S^{+} \approx \sqrt{N}c^{\dagger} .\label{l4a}
\end{eqnarray}
In this case, the Hamiltonian (\ref{l1a}) takes the form
\begin{eqnarray}
H &=& \, \Delta_{c} a^{\dagger}a +\Delta_{a}c^{\dagger}c +\omega_{m}b^{\dagger}b+g_{a}\left(c^{\dagger}a +a^{\dagger}c\right)  \nonumber \\
&+& g_{0}a^{\dagger}a \left(b^{\dagger}+b\right) +i\left(E_{L}a^{\dagger}-E^{\ast}_{L}a\right) ,\label{l1}
\end{eqnarray}
where $g_{a}=\sqrt{N}g_{e}$ is the coupling strength of the atomic mode to the cavity mode.

We take a strong driving field limit of $|E_{L}|\gg g_{0},g_{a}$, in which we can linearize the equations of motion for the bosonic operators of the modes around their semiclassical steady-state amplitudes~\cite{gm01,mg02}. Namely, we write the operators $c,b$ and $a$ as the sum of their steady-state amplitudes $\alpha_{a}, \alpha_{m}$ and $\alpha_{c}$, respectively, and small linear displacement (fluctuation) operators as
\begin{equation}
c\rightarrow\alpha_{a}+\delta c ,\quad b\rightarrow\alpha_{m}+\delta b,\quad a\rightarrow\alpha_{c}+\delta a ,\label{l2}
\end{equation}
where $\delta c, \delta b$ and $\delta a$ are the fluctuation operators.

Using the Hamiltonian (\ref{l1}) it is straightforward to derive the quantum Langevin equations for the fluctuation operators, which in a frame rotating with $\omega_{m}$ are
\begin{eqnarray}
\delta\dot{b} & =& -(\gamma_{m} +i\omega_{m})\delta b - ig_{m}(\delta a +\delta a^{\dagger}) -\sqrt{2\gamma_{m}}\, \delta b_{in} ,\nonumber \\
\delta\dot{a} & =& -(\kappa +i\Delta_{c})\delta a-ig_{a}\delta c-ig_{m}(\delta b +\delta b^{\dagger}) -\sqrt{2\kappa}\, \delta a_{in} ,\nonumber \\
\delta\dot{c} & =& -(\gamma_{a}+i\Delta_{a})\delta c -ig_{a}\delta a -\sqrt{2\gamma_{a}}\, \delta c_{in} ,\label{l3}
\end{eqnarray}
where $g_{m}=g_{0}|\alpha_{c}|$ is the effective mirror-cavity mode coupling. In writing~(\ref{l3}), we have included the damping rates of the modes and
the in-field operators $\delta b_{in}, \delta a_{in}$ and $\delta c_{in}$ describe the input noise to the modes.
The noise operators satisfy the Gaussian statistics with
\begin{equation}
\langle\delta a\rangle = \langle\delta b\rangle = \langle\delta c\rangle = 0 ,
\end{equation}
and nonzero correlations
\begin{eqnarray}
\langle\delta a_{in}(t)\delta a_{in}^{\dagger}(t^{\prime})\rangle &=& \langle\delta c_{in}(t)\delta c_{in}^{\dagger}(t^{\prime})\rangle = (n+1)\delta(t-t^{\prime}) ,\nonumber\\
\langle\delta a_{in}^{\dagger}(t)\delta a_{in}(t^{\prime})\rangle &=& \langle\delta c_{in}^{\dagger}(t)\delta c_{in}(t^{\prime})\rangle = n\delta(t-t^{\prime}) ,\nonumber\\
\langle\delta b_{in}(t)\delta b_{in}^{\dagger}(t^{\prime})\rangle &=& (n_{0}+1)\delta(t-t^{\prime}) ,\nonumber\\
\langle\delta b_{in}^{\dagger}(t)\delta b_{in}(t^{\prime})\rangle &=& n_{0}\delta(t-t^{\prime}) ,
\end{eqnarray}
where $n$ is the number of thermal excitations of the modes $a$ and $c$, and $n_{0}$ is the number of thermal excitations of the mode $b$.

It is seen from (\ref{l3}) that $\delta b$ oscillates at frequency $\omega_{m}$ while $\delta a$ and $\delta c$ oscillate at detunings $\Delta_{c}$ and $\Delta_{a}$, respectively.
We may introduce slowly varying variables and choose the detunings such that these new variables could be free of the oscillations. Namely, we may define slowly varying variables
\begin{eqnarray}
\delta b_{s} &=& \delta b e^{i\omega_{m}t} ,\quad \delta b^{s}_{in} = \delta b_{in} e^{i\omega_{m}t} ,\nonumber\\
\delta a_{s} &=& \delta a e^{-i\omega_{m}t} ,\quad \delta a^{s}_{in} = \delta a_{in} e^{-i\omega_{m}t} ,\nonumber\\
\delta c_{s} &=& \delta c e^{-i\omega_{m}t} ,\quad \delta c^{s}_{in} = \delta c_{in} e^{-i\omega_{m}t} ,
\end{eqnarray}
and in terms of these variables (\ref{l3}) becomes
\begin{eqnarray}
\delta\dot{b}_{s} & =& -\gamma_{m}\delta b_{s} - ig_{m}(\delta a_{s}e^{2i\omega_{m}t} +\delta a_{s}^{\dagger}) -\sqrt{2\gamma_{m}}\, \delta b^{s}_{in} ,\nonumber \\
\delta\dot{a}_{s} & =& -[\kappa +i(\Delta_{c}+\omega_{m})]\delta a_{s} -ig_{a}\delta c_{s} \nonumber\\
&& -ig_{m}(\delta b_{s}e^{-2i\omega_{m}t} +\delta b_{s}^{\dagger}) -\sqrt{2\kappa}\, \delta a^{s}_{in} ,\nonumber \\
\delta\dot{c}_{s} & =& -[\gamma_{a}+i(\Delta_{a}+\omega_{m})]\delta c_{s} -ig_{a}\delta a_{s} -\sqrt{2\gamma_{a}}\, \delta c^{s}_{in} .\label{l3u}
\end{eqnarray}
Choosing $\Delta_{c}=\Delta_{a}=-\omega_{m}$ and removing the terms oscillating at twice the frequency $\omega_{m}$, we obtain
\begin{eqnarray}
\delta\dot{b}_{s} & =& -\gamma_{m} \delta b_{s} - ig_{m}\delta a_{s}^{\dagger} -\sqrt{2\gamma_{m}}\, \delta b^{s}_{in} ,\nonumber \\
\delta\dot{a}_{s} & =& -\kappa \delta a_{s} -ig_{a}\delta c_{s} -ig_{m}\delta b_{s}^{\dagger}-\sqrt{2\kappa}\, \delta a^{s}_{in} ,\nonumber \\
\delta\dot{c}_{s} & =& -\gamma_{a}\delta c_{s} -ig_{a}\delta a_{s} -\sqrt{2\gamma_{a}}\, \delta c^{s}_{in} .\label{l3a}
\end{eqnarray}
Equation (\ref{l3a}) describes the dynamics of three coupled modes with a linear-type coupling between $a$ and $c$ modes, and a nonlinear-type coupling between the $a$ and $b$ modes.
Since the noise terms in (\ref{l3a}) satisfy the Gaussian statistics and the dynamics are linear, the steady-state of the system will be a Gaussian state.

Our purpose in this paper is to investigate quantum steering between two modes of this three mode system. Thus, we will adiabatically eliminate one of the modes from the dynamics by assuming that the mode is damped with a rate which is much larger than damping rates of the remaining modes. Thus, the fast damped mode can be treated as an "auxiliary" mode to the other modes. We will consider two cases. In the first, we will assume that the atomic mode $c$ is rapidly damped that the mode can be treated as an "auxiliary" mode to the cavity mode. In the second, we will assume that the cavity mode $a$ is rapidly damped that the mode can appear as an "auxiliary" mode to both atomic and mirror modes.

Before proceeding further, we briefly discuss where the assumption of fast damped modes may be satisfied in currently available experimental systems. Since the atomic damping rate $\gamma_{a}$ is proportional to the square of the absolute value of the atomic transition dipole moment, $|\vec{\mu}_{a}|^{2}$, the condition of a fast damped atomic mode can be realised simply by choosing atoms with a large $\vec{\mu}_{a}$. The situation of the fast damped cavity mode can be met in recently experimentally realized optomechanical systems~\cite{pp13,pp16}, in which damping parameters were $\kappa/2\pi = 2.6$ MHz and $\gamma_{m}/2\pi =0.18$ Hz. Clearly, the condition of $\kappa\gg \gamma_{m}$ could be well satisfied in these experiments.

\subsection{The case of fast damped atomic mode}

Our first example considers the atomic mode an auxiliary mode to the parametrically-type coupled cavity and mirror modes. We adiabatically eliminate the atomic modes assuming that the damping rate of the mode $\gamma_{a}$ is much larger than the damping rates of the other modes. In this case, we formally integrate the equation of motion $\delta\dot{c}_{s}$, appearing in (\ref{l3a}), and get
\begin{eqnarray}
\delta c_{s} &=& \delta c_{s}(0)e^{-\gamma_{a}t} -ig_{a}e^{-\gamma_{a}t}\int_{0}^{t}dt^{\prime}\delta a_{s}(t^{\prime})e^{\gamma_{a}t^{\prime}} \nonumber\\
&-&\sqrt{2\gamma_{a}}\int_{0}^{t}dt^{\prime}\delta c^{s}_{in}(t^{\prime})e^{-\gamma_{a}(t-t^{\prime})} .\label{csa}
\end{eqnarray}
When the atomic mode is fast damped we have $\gamma_{a}t\gg 1$, even at short times. In this case, we can make the adiabatic approximation that over such short times the operators of the other modes do not change much, so that we can approximate $a_{s}(t^{\prime})\approx a_{s}(t)$. Then, in the limit of $\gamma_{a}t\gg 1$ the solution for $\delta c_{s}$ reduces to
\begin{eqnarray}
\delta c_{s} \approx -i\frac{g_{a}}{\gamma_{a}}\delta a_{s} -\sqrt{2\gamma_{a}}\int_{0}^{t}dt^{\prime}\delta c^{s}_{in}(t^{\prime})e^{-\gamma_{a}(t-t^{\prime})} .\label{l4}
\end{eqnarray}
When this result is inserted into the equation of motion $\delta\dot{a}$, the set of coupled differential equations (\ref{l3}) reduces to the following set of two coupled equations of motion for the cavity and mirror modes modified by the presence of the auxiliary atomic mode
\begin{eqnarray}
\delta\dot{a}_{s} & =& -\left(\kappa +C_{a}\right)\delta a_{s} -ig_{m}\delta b_{s}^{\dagger} +i\sqrt{2C_{a}}\delta\tilde{c}^{s}_{in} -\sqrt{2\kappa}\,\delta a^{s}_{in} ,\nonumber\\
\delta\dot{b}_{s} & =& -\gamma_{m} \delta b_{s} - ig_{m}\delta a_{s}^{\dagger} -\sqrt{2\gamma_{m}}\,\delta b^{s}_{in} ,\label{l5}
\end{eqnarray}
where $C_{a}=g_{a}^{2}/\gamma_{a}$, and
\begin{eqnarray}
\delta\tilde{c}^{s}_{in} = \gamma_{a}\int_{0}^{t}dt^{\prime}\delta c^{s}_{in}(t^{\prime})e^{-\gamma_{a}(t-t^{\prime})} .\label{l6}
\end{eqnarray}

For each mode we introduce the quadrature phase amplitudes $X_{u}=(\delta u +\delta u^{\dagger})/\sqrt{2}$ and $P_{u}=(\delta u -\delta u^{\dagger})/\sqrt{2}i$, $(\delta u=\delta a_{s},\delta b_{s},\delta c_{s})$ and by solving~(\ref{l5}) for the steady-state, we will determine the variances of the quadrature amplitudes of the modes and correlations between the modes.
It is straightforward to show that the steady-state variances of the quadrature phase amplitudes of the cavity and mirror modes and non-zero correlation functions are
\begin{eqnarray}
\Delta^{2} X_{a} = \Delta^{2} P_{a} &=& \left(n+\frac{1}{2}\right) \nonumber\\
&+&\frac{1}{2}\frac{\left(n+n_{0}+1\right)\gamma g_{m}^{2}}{\left(\gamma +\frac{1}{2}C_{a}\right)\!\left[\gamma\left(\gamma +C_{a}\right) -g_{m}^{2}\right]} ,\nonumber\\
\Delta^{2} X_{b} = \Delta^{2} P_{b} &=& \left(n_{0}+\frac{1}{2}\right) \nonumber\\
&+&\frac{1}{2}\frac{\left(n+n_{0}+1\right)\left(\gamma+C_{a}\right)g_{m}^{2}}{\left(\gamma +\frac{1}{2}C_{a}\right)\!\left[\gamma\left(\gamma +C_{a}\right) -g_{m}^{2}\right]}  ,\nonumber\\
\langle X_{a}P_{b}\rangle = \langle X_{b}P_{a}\rangle &=& -\frac{1}{2}\frac{\left(n+n_{0}+1\right)\gamma(\gamma +C_{a})g_{m}}{\left(\gamma +\frac{1}{2}C_{a}\right)\left[\gamma\left(\gamma +C_{a}\right)-g_{m}^{2}\right]} ,\label{l7}
\end{eqnarray}
where $\Delta^{2}U =\langle U^{2}\rangle -\langle U\rangle^{2}$ and $\gamma\equiv\gamma_{m}=\kappa$. Solving~(\ref{l4}) for the steady-state, we readily find that $\Delta^{2} X_{c} = \Delta^{2} P_{c}=(n+1/2)$, i.e., the fluctuations of the fast damped atomic mode are not affected by the other modes. There is no back action of the modes on the fast damped mode. Consequently, we can call the atomic mode as an auxiliary mode.

Equation (\ref{l7}) shows that the effect of the auxiliary mode is to introduce an asymmetry in the variances of the quadrature components of the modes. It is not difficult to see that in the case of $C_{a}\neq 0$ and $n=n_{0}$ we have $\Delta^{2} X_{a} < \Delta^{2} X_{b}$, i.e., the fluctuations of the mode to which the auxiliary mode is directly coupled are smaller than the fluctuations of the other mode. However, the correlations between the modes are present even in the absence of the auxiliary mode, $C_{a}=0$.

\subsection{The case of fast damped cavity mode}

For our second example, we consider the steady-state properties of the indirectly coupled atomic and mirror modes interacting with the rapidly damped cavity mode, $\kappa\gg \gamma_{a},\gamma_{m}, g_{a}, g_{m}$. Now, we formally integrate the equation of motion for $\delta a_{s}$, appearing in~(\ref{l3a}), and make the adiabatic approximation. Then, substituting the resulting expression into the equations of motion $\delta \dot{b}_{s}$ and $\delta \dot{c}_{s}$, we get
\begin{eqnarray}
\delta\dot{b}_{s} &=& -(\gamma_{m} -G)\delta b_{s} +\sqrt{GG_{a}}\delta c_{s}^{\dagger} + i\sqrt{2G}\delta\tilde{a}_{in}^{s\dagger} -\sqrt{2\gamma_{m}}\, \delta b^{s}_{in} ,\nonumber \\
\delta\dot{c}_{s} &=& -(\gamma_{a}+G_{a})\delta c_{s} -\sqrt{GG_{a}}\delta b_{s}^{\dagger} + i\sqrt{2G_{a}}\delta\tilde{a}^{s}_{in} -\sqrt{2\gamma_{a}}\, \delta c^{s}_{in} ,\label{l8}
\end{eqnarray}
where $G=g_{m}^{2}/\kappa$, $G_{a}=g_{a}^{2}/\kappa$, and
\begin{eqnarray}
\delta\tilde{a}^{s}_{in} = \kappa \int_{0}^{t}dt^{\prime}a^{s}_{in}(t^{\prime})e^{-\kappa (t-t^{\prime})} .\label{l9}
\end{eqnarray}

Solving~(\ref{l8}) for the steady-state, we arrive at the following expressions for the variances and correlations
\begin{eqnarray}
\Delta^{2} X_{b} &=& \Delta^{2} P_{b} = \left(n_{0} +\frac{1}{2}\right) +\left(n+n_{0}+1\right)\frac{2G\left(\gamma_{G}+\frac{1}{2}G\right)}{\gamma_{G}\left(\gamma +\gamma_{G}\right)} ,\nonumber\\
\Delta^{2} X_{c} &=& \Delta^{2} P_{c} = \left(n+ \frac{1}{2}\right) +\left(n+n_{0}+1\right)\frac{GG_{a}}{\gamma_{G}\left(\gamma +\gamma_{G}\right)} ,\nonumber\\
\langle X_{b}X_{c}\rangle &=& -\langle P_{b}P_{c}\rangle = -\left(n+n_{0}+1\right)\frac{\sqrt{GG_{a}}\left(\gamma_{G}+G\right)}{\gamma_{G}\left(\gamma +\gamma_{G}\right)} ,\label{l10}
\end{eqnarray}
where $\gamma_{G} =\gamma-(G-G_{a}) $. It can be shown that the variances of the quadrature components of the fast damped mode are not affected by the other modes, $\Delta^{2} X_{a} = \Delta^{2} P_{a}=(n+1/2)$.
Notice an asymmetry in the variances that for $n=n_{0}$, $\Delta^{2} X_{c} < \Delta^{2} X_{b}$. Again, we see that the effect of the auxiliary mode is to introduce an asymmetry between the variances of the quadrature components of the modes. Moreover, the correlations between the modes are different from zero only if $G\neq 0$\, (i.e., $g_{m}\neq 0)$.

It is worth noting that the variances $\Delta^{2} X_{i}$ and $\Delta^{2} P_{i}\, (i=a,b,c)$ as given in Equations~(\ref{l7}) and (\ref{l10}), are not reduced below the quantum limit, i.e. $\Delta^{2} X_{i}>1/2$ and $\Delta^{2} P_{i}>1/2$. However, the presence of nonzero correlations between the modes suggests that the variances of linear combinations of the modes could be reduced below the quantum limit. For example, in the case of rapidly damped atomic mode the variances in linear combinations, $X_{a}+hP_{b}$ and $P_{a}+hX_{b}$ could be reduced below the quantum limit. Similarly, in the case of rapidly damped cavity mode the variances in linear combinations, $X_{a}+hX_{b}$ and $P_{a}-hP_{a}$ could also be reduced below the quantum limit. Here, $h$ is a parameter which can be chosen to minimize the variance relative to the quantum limit.
We point out that the variances of such linear combinations are involved in the criteria for entanglement and quantum steering.

\section{Criteria for bipartite steering and entanglement}\label{Sec3}

Since the steady-states (\ref{l7}) and (\ref{l10}) of our system are two-mode Gaussian states, we will adopt the Reid criterion to determine conditions for quantum steering~\cite{md89} (see also~\cite{cj09,rd09}). This is justified because the criterion is applicable to non-idealized Gaussian continuous variable systems, i.e. for decohered mixed Gaussian states. The Reid criterion is based on an accuracy of inference defined as the root mean square of the variances, $\Delta_{{\rm inf},j}X_{i}\equiv \sqrt{\Delta_{{\rm inf},j}^{2}X_{i}}$ and $\Delta_{{\rm inf},j}P_{i}\equiv \sqrt{\Delta_{{\rm inf},j}^{2}P_{i}}$
of the conditional distributions for a result of measurement of the quadratures $X_{i}$ and $P_{i}$ of the mode $i$, based on the results of the measurement of quadratures of the mode $j$. The inferred variance $\Delta_{{\rm inf},j}X_{i}^{{\rm out}}$ can be written as~\cite{md89,cj09,rd09,rt15}
\begin{equation}
\Delta_{{\rm inf},j}X_{i} =\Delta\left(X_{i} +h_{j}O_{j}\right), \quad \Delta_{{\rm inf},j}P_{i} =\Delta\left(P_{i} +h^{\prime}_{j}O^{\prime}_{j}\right) ,\label{equ1}
\end{equation}
where the quadrature $O_{j}(O^{\prime}_{j})$ is selected either $O_{j}(O^{\prime}_{j})\equiv X_{j}$ or $O_{j}(O^{\prime}_{j})\equiv P_{j}$, depending on the type of the correlations between the modes, i.e., depending on whether nonzero correlations between the modes are of the $X-X$, $P-P$ or $X-P$ type. The parameters $h_{j}$ and $h^{\prime}_{j}$ are constants chosen such that they minimise the variances and thus the uncertainty product
\begin{equation}
 E_{i|j} \equiv \Delta_{{\rm inf},j}X_{i}\Delta_{{\rm inf},j}P_{i} = \Delta(X_{i}+h_{j}O_{j})\Delta(P_{i}+h^{\prime}_{j}O^{\prime}_{j}) .\label{l21a}
\end{equation}
The values of $h_{j}$ and $h_{j}^{\prime}$ which minimise the variances are found by taking derivatives of the variances, respectively, over $h_{j}$ and $h_{j}^{\prime}$, and setting the derivatives to zero we find that the variances minimise when
\begin{equation}
h_{j}= -\frac{\left\langle X_{i}O_{j}\right\rangle }{\Delta^{2}O_{j}} ,\quad h^{\prime}_{j}= -\frac{\left\langle P_{i}O^{\prime}_{j}\right\rangle }{\Delta^{2}O^{\prime}_{j}} .\label{l14a}
\end{equation}
This results in the following expression for $E_{i|j}$ corresponding to the optimal variances
\begin{eqnarray}
E_{i|j} = \sqrt{\Delta^{2}X_{i}\left[1 -\left(C_{X_{i},O_{j}}\right)^{2}\right]}\sqrt{\Delta^{2}P_{i}\left[1 -\left(C_{P_{i},O^{\prime}_{j}}\right)^{2}\right]} ,\label{l15a}
\end{eqnarray}
where
\begin{equation}
C_{U_{i},W_{j}} = \frac{\left\langle U_{i}W_{j}\right\rangle}{\sqrt{\Delta^{2}U_{i}\Delta^{2}W_{j}}} ,\quad (U_{i}=X_{i},P_{i}; W_{j} =O_{j},O^{\prime}_{j})
\end{equation}
is the correlation coefficient for the quadrature components.

We say that the mode $i$ is steered by the mode $j$ whenever $E_{i|j} < 1/2$. Because of the asymmetry property of the steering parameter $E_{i|j}$, the reverse steering of the mode $j$ by the mode $i$ is not always true, that $E_{i|j}<1/2$ does not necessary mean that $E_{j|i}<1/2$. The situation of $E_{i|j}<1/2$ and $E_{j|i}>1/2$ is referred to as the {\it one-way steering} that mode $i$ is steered to entanglement by the mode $j$, but the mode $j$ is not steered by mode $i$ to entanglement. The situation of $E_{i|j}<1/2$ and $E_{j|i}<1/2$ is referred to as a {\it two-way steering}. In this case, both mode $i$ steers mode $j$ and in the same time mode $j$ steers mode $i$ to entanglement.

Variances of linear combinations of the quadrature components are also involved in the criteria for entanglement between modes. For example, in the case of $X-P$ type correlations between the modes, the asymmetric Duan-Simon criterion for entanglement is defined as~\cite{hr13,dg00,rs00}
\begin{eqnarray}
\Delta_{i,j}^{h} = \frac{\Delta^{2}\!\left(X_{i}+hP_{j}\right) +\Delta^{2}\left(P_{i}+hX_{j}\right)}{1+h^{2}} .\label{l20}
\end{eqnarray}
The modes are entangled if the condition $\Delta_{i,j}^{h}<1$ is satisfied.

A two-mode Gaussian state can be characterised by its symmetrically ordered correlation matrix $V$ with components $V_{ij} = \langle O_{i}O_{j}+O_{j}O_{i}\rangle/2$, where $O_{i},O_{j}$ are components of the vector $\vec{O}^{T}= (X_{l},P_{l},X_{k},P_{k})$ composed of the quadrature phase amplitudes of the two modes $l$ and $k$. Then the entanglement of the state can be quantified by means of the logarithmic negativity $E_{{\cal N}}$ defined as
\begin{equation}
E_{{\cal N}} = {\rm max}[0,-\ln(2\lambda)] ,
\end{equation}
where $\lambda = 2^{-1/2}\{\Sigma(V)-[\Sigma(V)^{2}-4{\rm det}(V)]^{1/2}\}^{1/2}$, with $\Sigma(V)={\rm det}(V_{l})+{\rm det}(V_{k})-2{\rm det}(V_{{\rm corr}})$ given in terms of $2\times 2$ block matrices of the correlation matrix
\begin{eqnarray}
V &=& \left(\begin{array}{cc}
V_{l}&V_{{\rm corr}}\\
V^{T}_{{\rm corr}}&V_{k}
\end{array}\right) .
\end{eqnarray}
The $2\times 2$ matrices are of the form
\begin{eqnarray}
V_{l} &=& \left(\begin{array}{cc}
\langle X_{l}^{2}\rangle&\langle X_{l}P_{l}\rangle\\
\langle P_{l}X_{l}\rangle&\langle P_{l}^{2}\rangle
\end{array}\right) , V_{k} = \left(\begin{array}{cc}
\langle X_{k}^{2}\rangle&\langle X_{k}P_{k}\rangle\\
\langle P_{k}X_{k}\rangle&\langle P_{k}^{2}\rangle
\end{array}\right) ,
\end{eqnarray}
and
\begin{eqnarray}
V_{{\rm corr}} &=& \left(\begin{array}{cc}
\langle X_{l}X_{k}\rangle&\langle X_{l}P_{k}\rangle\\
\langle P_{l}X_{k}\rangle&\langle P_{l}P_{k}\rangle
\end{array}\right) .
\end{eqnarray}

The above discussed criteria show that correlations between the modes are necessary for entanglement and quantum steering and should be strong enough to satisfy the required inequalities. However, for quantum steering to be observed the correlations between modes must be stronger than those for entanglement. Specifically, for the case determined by the steady-state (\ref{l7}), the criterion (\ref{l15a}) for quantum steering between the cavity and mirror modes yields the inequality
\begin{equation}
|\langle X_{a}P_{b}\rangle| > \sqrt{\Delta^{2} P_{b}\left(\Delta^{2} X_{a} -1/2\right)} ,\label{l30a}
\end{equation}
whereas the criterion for entanglement $(\lambda <1/2)$ yields
\begin{equation}
|\langle X_{a}P_{b}\rangle| > \sqrt{\left(\Delta^{2} P_{b}-1/2\right)\left(\Delta^{2} X_{a} -1/2\right)} .\label{l31a}
\end{equation}
Clearly, the inequality for the correlations to observe quantum steering is stronger than that required for entanglement.

\begin{figure}[ht]
\includegraphics[width=0.5\columnwidth]{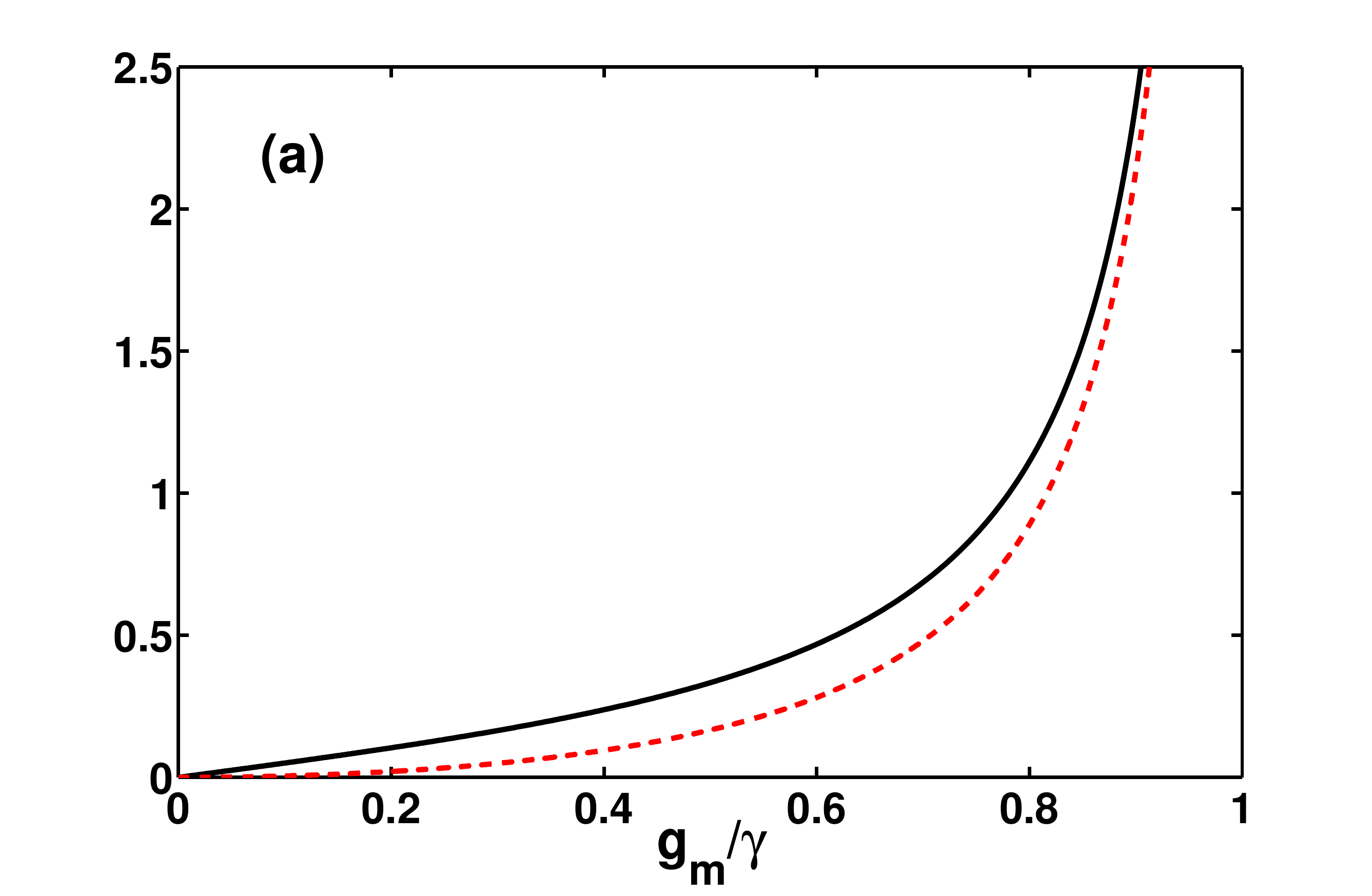}
\includegraphics[width=0.5\columnwidth]{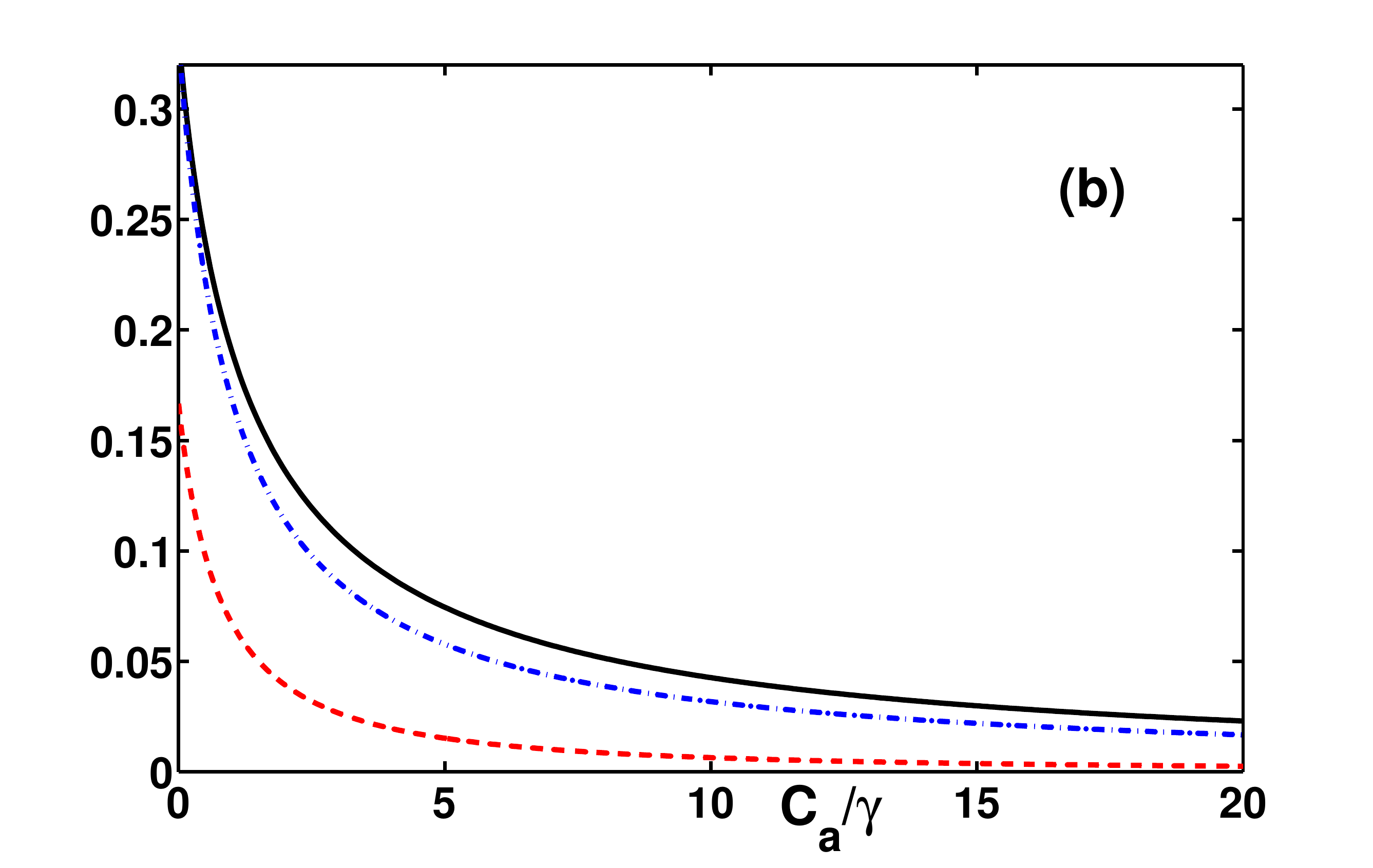}
\caption{(a) Correlation function $|\langle X_{a}P_{b}\rangle|$ (solid black line) and threshold for entanglement $\sqrt{\left(\Delta^{2} P_{b}-1/2\right)\left(\Delta^{2} X_{a} -1/2\right)}$ (dashed red line) plotted as a function of $g_{m}$ for $C_{a}=0$ and $n=n_{0}=0$. For these parameter values the correlation function is equal to the threshold value for quantum steering. (b) Correlation function $|\langle X_{a}P_{b}\rangle|$ (solid black line) and thresholds for entanglement (dashed red line) and quantum steering (dashed-dotted blue line) plotted as a function of $C_{a}$ for $g_{m}=\gamma/2$ and $n=n_{0}=0$.}
\label{fig2a}
\end{figure}

Figure~\ref{fig2a} shows the correlation function $|\langle X_{a}P_{b}\rangle|$ and the thresholds for steering and entanglement calculated from (\ref{l30a}) and (\ref{l31a}) using the steady-state solutions (\ref{l7}). In the absence of the auxiliary mode $(C_{a}=0)$ the correlations are stronger than the threshold for entanglement but are not stronger than the threshold for quantum steering. For the parameter values used the correlations are equal to the threshold for quantum steering over the entire range of $g_{m}$. Figure~\ref{fig2a}(b) shows the effect of the auxiliary mode $(C_{a}\neq 0)$ on the correlations and the thresholds for steering and entanglement. It is seen that the correlations and the thresholds decrease with the increasing $C_{a}$. However, the correlations decrease slower than the thresholds leading to the correlations larger than the threshold for quantum steering.
Note that threshold for entanglement is always less than threshold for quantum steering.

\section{Role of an auxiliary mode in the creation of bipartite steering}\label{Sec4}

We now proceed to discuss in details the effect of an auxiliary mode on the bipartite quantum steering properties of two modes to which the auxiliary mode is coupled.

\subsection{Atomic mode as an auxiliary mode}

Let us first consider the criterion for bipartite steering between the cavity and mirror modes with the atomic mode appearing as an auxiliary mode.
Note that in this case the auxiliary mode is directly coupled to the cavity mode, and the cavity and mirror modes are coupled by the parametric-type interaction.

According to~(\ref{l7}), there are nonzero $\langle X_{a}P_{b}\rangle$ and $\langle X_{b}P_{a}\rangle$ correlations between the modes. Therefore, in the definition of the steering parameter~(\ref{l15a}), we choose $O_{b}\equiv P_{b}$ for the quadrature of the mode $b$. In addition, since $\Delta^{2} P_{a}=\Delta^{2} X_{a}$, $\Delta^{2} P_{b}=\Delta^{2} X_{b}$ and $\langle X_{a}P_{b}\rangle=\langle X_{b}P_{a}\rangle$, we see that the steering parameter $E_{a|b}$ is of the form
\begin{eqnarray}
E_{a|b} = \Delta^{2}X_{a} -\frac{\left\langle X_{a}P_{b}\right\rangle^{2} }{\Delta^{2}P_{b}} .\label{l15}
\end{eqnarray}

Similarly, we can find that the steering parameter $E_{b|a}$ describing steering of the mirror mode $b$ by the cavity mode $a$ is of the form
\begin{eqnarray}
E_{b|a} = \Delta^{2}X_{b} -\frac{\left\langle X_{a}P_{b}\right\rangle^{2} }{\Delta^{2}P_{a}} .\label{l16}
\end{eqnarray}

Since $\Delta^{2}X_{a}\geq \frac{1}{2}$ and $\Delta^{2}X_{b}\geq \frac{1}{2}$, we see that the minimum requirement for steering to be possible is that the modes $a$ and $b$ are correlated, i.e., that $\langle X_{a}P_{b}\rangle\neq 0$. Note that the requirement that the modes should be correlated is necessary but not sufficient for quantum steering. The correlation $\langle X_{a}P_{b}\rangle$ although nonzero may not be large enough to reduce $E_{a|b}$ and/or $E_{b|a}$ below the threshold value $1/2$. For quantum steering to occur, $\langle X_{a}P_{b}\rangle$ should be sufficiently large to enforce the inequality $E_{a|b}<1/2$ and/or $E_{b|a}<1/2$.

The variances and the correlation function, which are needed in~(\ref{l15}) and (\ref{l16}) have already been obtained and are given in~(\ref{l7}). To illustrate the analytic structure of the steering parameters, we assume equal thermal excitations of the modes, $n=n_{0}$. Then, we find
\begin{eqnarray}
&E_{a|b} =\left(n_{0}+\frac{1}{2}\right) \nonumber\\
&\times\left\{1 -\frac{\gamma C_{a}g_{m}^{2}}{\left(\gamma +\frac{1}{2}C_{a}\right)\left[2\gamma\!\left(\gamma +\frac{1}{2}C_{a}\right)\!\left(\gamma +C_{a}\right)+C_{a}g_{m}^{2}\right]} \right\} ,\label{l17}
\end{eqnarray}
and
\begin{eqnarray}
&E_{b|a} =\left(n_{0}+\frac{1}{2}\right)\nonumber\\
&\times\left\{1 +\frac{C_{a}\left(\gamma +C_{a}\right)g_{m}^{2}}{\left(\gamma +\frac{1}{2}C_{a}\right)\!\left[2\gamma\left(\gamma +\frac{1}{2}C_{a}\right)\left(\gamma +C_{a}\right)-C_{a}g_{m}^{2}\right]} \right\} .\label{l18}
\end{eqnarray}
We see from~(\ref{l17}) and (\ref{l18}) that $E_{a|b}$ can be reduced below threshold for quantum steering, but $E_{b|a}$ is always larger than the threshold value $1/2$.
The results given by~(\ref{l17}) and (\ref{l18}) clearly show that in the presence of the auxiliary mode coupled to the cavity mode, $C_{a}\neq 0$ and $n_{0}=0$, the cavity mode $a$ is always steered by the mirror mode $b$. This means once the auxiliary atomic mode is coupled to the cavity mode, one-way steering, $E_{a|b}<1/2$, can be observed. If the auxiliary atomic mode was decoupled from the cavity mode $(C_{a}=0)$, then the steering parameters would be
\begin{equation}
E_{a|b}=E_{b|a} =\left(n_{0}+\frac{1}{2}\right) ,\label{l19}
\end{equation}
which are constant and not reduced below the threshold $1/2$. This clearly shows that the fast damped auxiliary atomic mode is responsible for steering of the cavity mode $a$ by the mirror mode $b$.

\begin{figure}[ht]
\includegraphics[width=0.5\columnwidth]{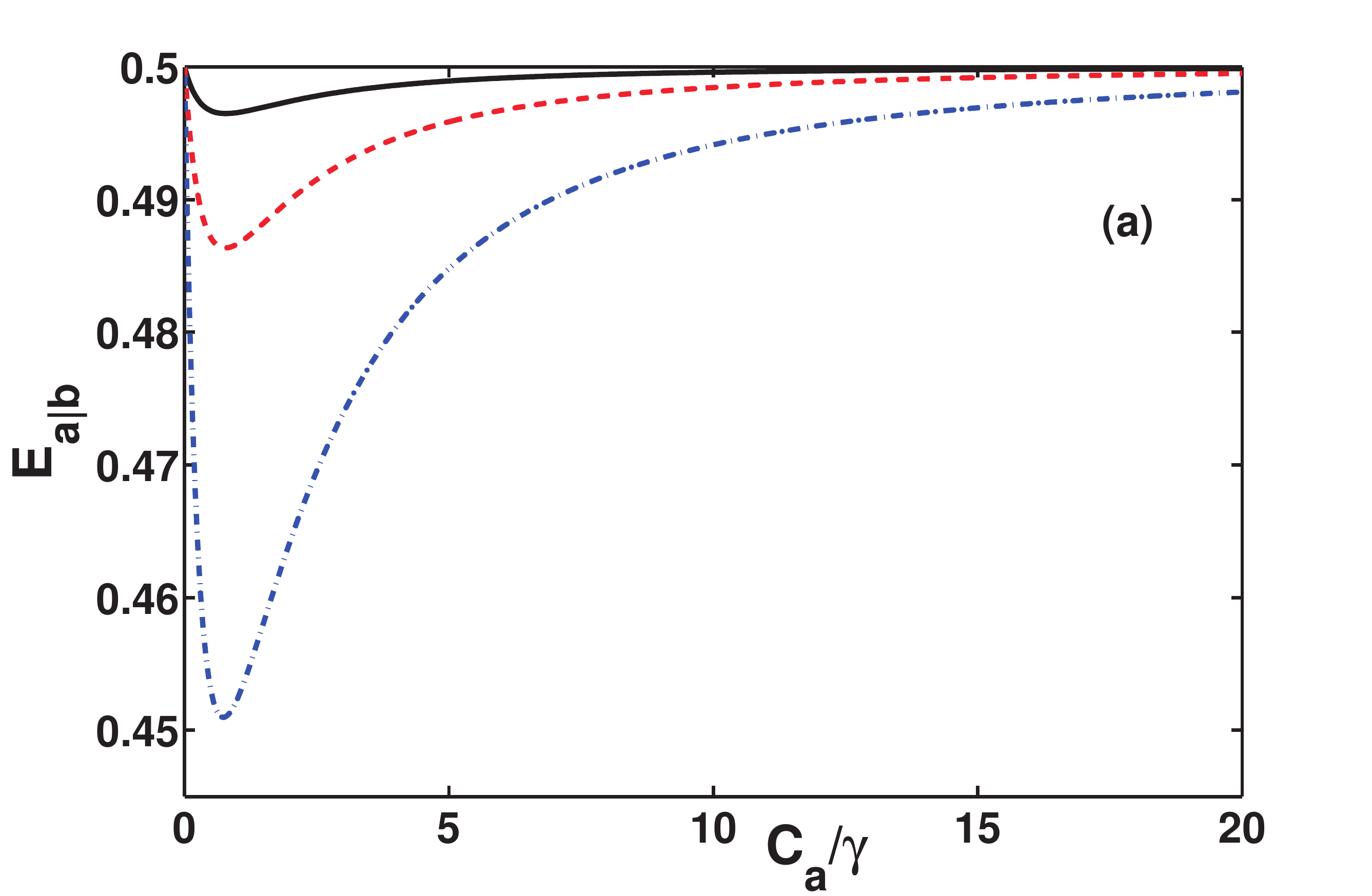}
\includegraphics[width=0.5\columnwidth]{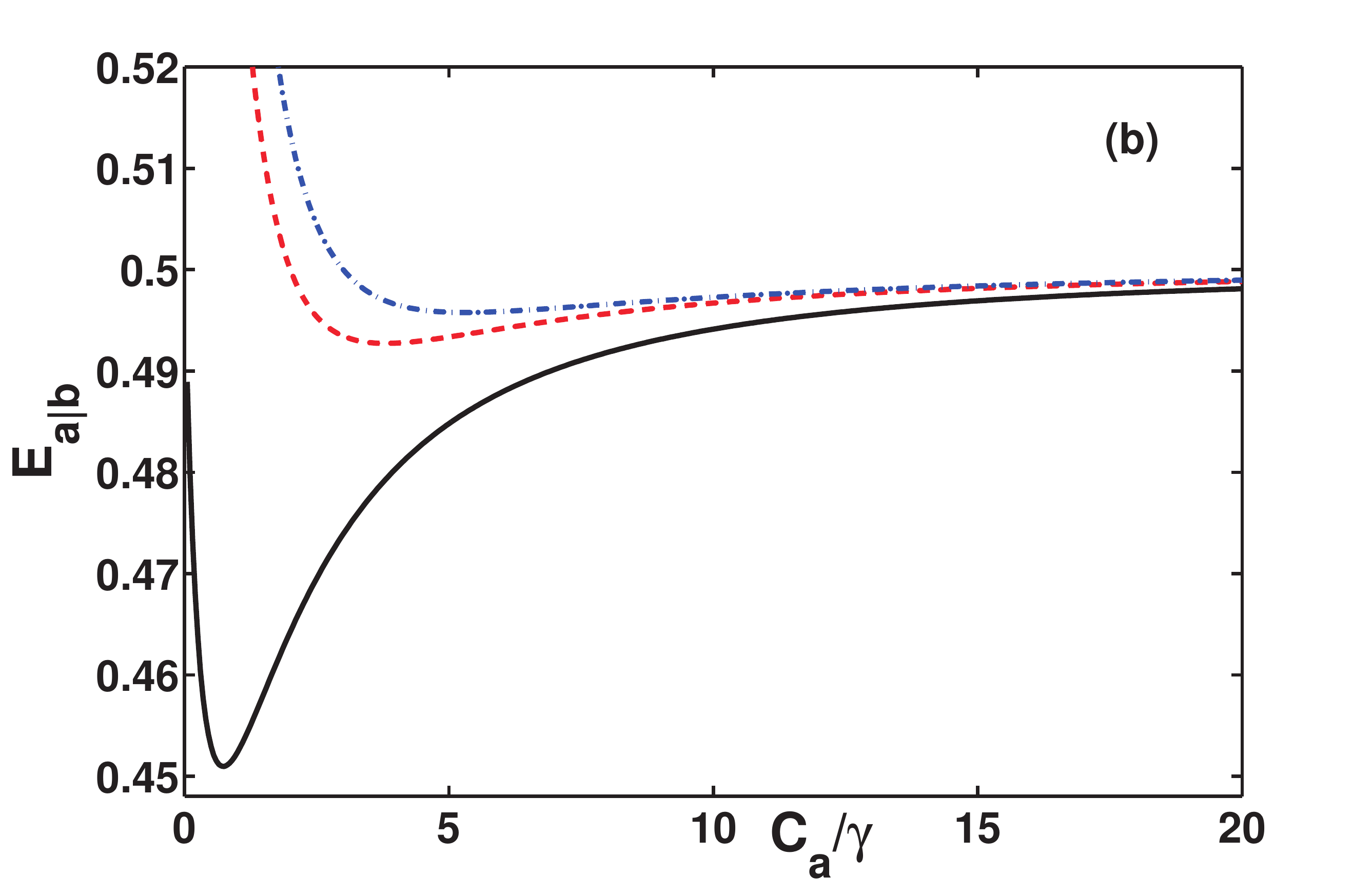}
\caption{(a) Illustration of the effect of the coupling constant $g_{m}$ on the variation of the steering parameter $E_{a|b}$ with $C_{a}$ in the case no thermal excitation at the modes $(n=n_{0}=0)$. Solid black line corresponds to $g_{m}=0.25\gamma$, dashed red line to $g_{m}=0.5\gamma$, and dotted-dashed blue line to $g_{m}=\gamma$. (b) Illustration of the effect of the thermal excitation at the mirror mode on the variation of $E_{a|b}$ with $C_{a}$ in the case no thermal excitation at the cavity mode $(n=0)$ and $g_{m}=\gamma$. Solid black line corresponds to $n_{0}=0$, dashed red line to $n_{0}=1$, and dotted-dashed blue line to $n_{0}=1.5$. }
\label{fig2}
\end{figure}

Figure~\ref{fig2} shows the steering parameter $E_{a|b}$ as a function of the coupling strength $C_{a}$ of the auxiliary atomic mode to the cavity mode. In the absence of the thermal excitation at the modes, $n=n_{0}=0$, one-way steering of the mode $a$ by the mode $b$ is present over the entire range of $C_{a}$ and the largest steering occurs for $C_{a}\approx \gamma$. As $C_{a}$ increases, the steering decays steadily. The important feature however is that the steering is never destroyed completely. In the presence of the thermal excitation at the mirror mode, the steering is lost at small $C_{a}$, but is almost insensitive to the thermal excitation at large $C_{a}$.

The physical reason for the occurrence of one-way steering with no possibility for two-way steering is in the asymmetry of the variances of the quadrature components of the modes.  The auxiliary mode affects the variances of the quadrature components of the modes such that $\Delta^{2}X_{a}<\Delta^{2}X_{b}$. This indicates that only a mode of larger fluctuations can steer the mode of smaller fluctuations. Thus, the fluctuations act directionally, to prevent a mode to be steered whose the fluctuations are larger than that of the steering mode.

\subsection{Cavity mode as an auxiliary mode}

Consider now the case where the cavity mode is fast damped so it can be treated as an auxiliary mode to the atomic and mirror modes. Note that in this case the auxiliary mode is coupled to both the atomic and mirror modes. Thus, the  auxiliary mode simultaneously affects both modes.

In this case, the steering parameter $E_{c|b}$ determining quantum steering of the atomic mode $c$ by the mirror mode $b$ is given by
\begin{eqnarray}
E_{c|b} & =& \Delta^{2}X_{c} -\frac{\left\langle X_{c}X_{b}\right\rangle^{2} }{\Delta^{2}X_{b}} .\label{l26}
\end{eqnarray}
Similarly, we can show that the minimized steering parameter $E_{b|c}$ is of the form
\begin{eqnarray}
E_{b|c} & =& \Delta^{2}X_{b} -\frac{\left\langle X_{c}X_{b}\right\rangle^{2} }{\Delta^{2}X_{c}} .\label{l27}
\end{eqnarray}

To evaluate the steering parameters we will use the results for the variances and the correlation function given in~(\ref{l10}). In order to get some inside into the analytic structure of the steering parameters, we put $n=n_{0}$ and then we obtain the following analytical result for $E_{c|b}$,
\begin{eqnarray}
E_{c|b} &=& \left(n_{0}+\frac{1}{2}\right) \nonumber\\
&\times& \left\{1 -\frac{2GG_{a}\left(G_{a}-G\right)}{\left(\gamma +\gamma_{G}\right)\left[\gamma_{G}\left(\gamma +\gamma_{G}\right)+4G\left(\gamma_{G}+\frac{1}{2}G\right)\right]} \right\} ,\label{l28}
\end{eqnarray}
For $E_{b|c}$ we find
\begin{eqnarray}
E_{b|c} &=&\left(n_{0}+\frac{1}{2}\right)\nonumber\\
&\times& \left\{1+\frac{2G\left[4\gamma\gamma_{G} -G\left(G_{a}-G\right)\right]}{\left(\gamma +\gamma_{G}\right)\left[\gamma_{G}\left(\gamma +\gamma_{G}\right)+2GG_{a}\right]}\right\} .\label{l29}
\end{eqnarray}
Equation (\ref{l28}) shows that in order to have $E_{c|b}<1/2$, one must have $G_{a}>G$. In other words, the coupling strength of the auxiliary cavity mode to the atomic mode should be larger than the coupling strength to the mirror mode to observe quantum steering between the modes $c$ and $b$. Thus, some kind of an asymmetry is required to be present in the coupling of the auxiliary mode to the two other modes.

Equation (\ref{l29}) shows that in order to have $E_{b|c}<1/2$, one must have $G_{a}>G$ and $4\gamma\gamma_{G}<G(G_{a}-G)$.
Hence, there are two regimes for the observation of one-way and two-way quantum steering. For $G_{a}>G$ and $4\gamma\gamma_{G}>G(G_{a}-G)$, the only steering possible is one-way steering $E_{c|b}<\frac{1}{2}$ and $E_{b|c}>\frac{1}{2}$. For $G_{a}>G$ and $4\gamma\gamma_{G}<G(G_{a}-G)$, both $E_{c|b}$ and $E_{b|c}$ can be reduced below $1/2$. In this case, two-way steering becomes possible. The involvement of $\gamma$ indicates that there is a spontaneous emission (losses) dependent minimal coupling strength $G_{a}$ required to observe the two-way steering.
\begin{figure}[ht]
\includegraphics[width=0.5\columnwidth]{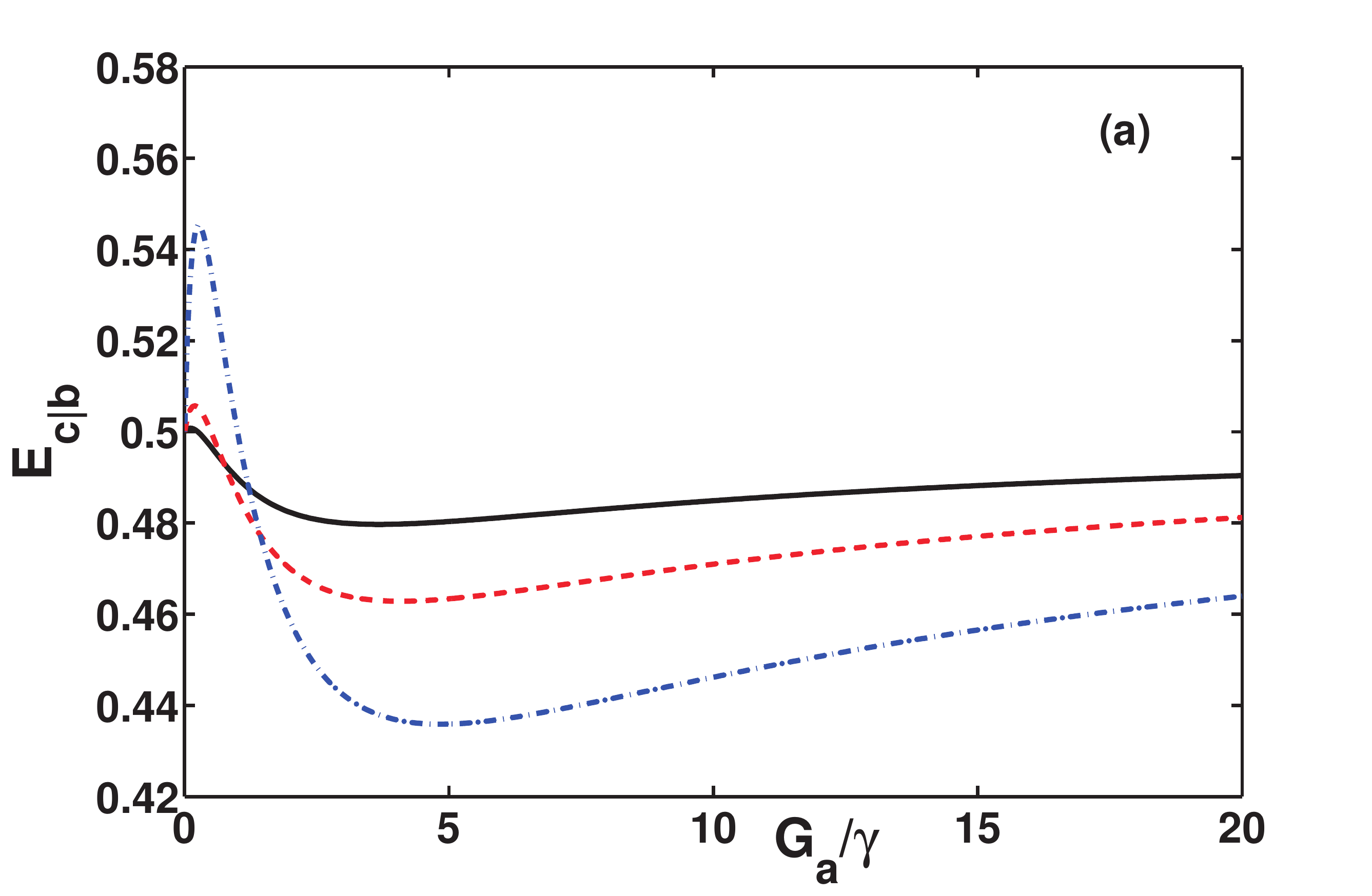}
\includegraphics[width=0.5\columnwidth]{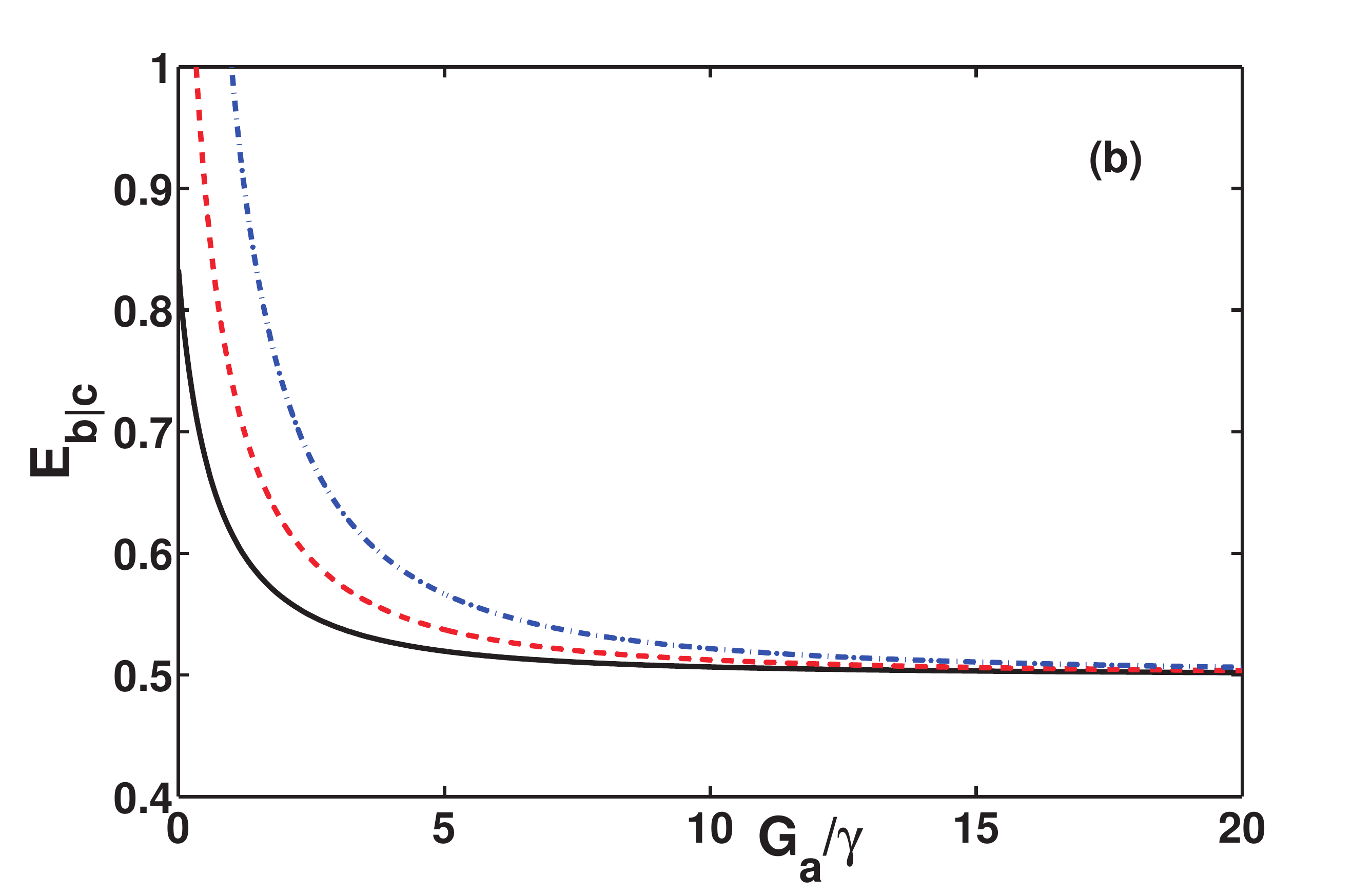}
\caption{The variation of the steering parameters (a) $E_{c|b}$ and (b) $E_{b|c}$ with $G_{a}/\gamma$ for $n=n_{0}=0$ and for several different values of the coupling constant $G\leq \gamma$: $G=0.25\gamma$ (solid black line), $G=0.5\gamma$ (dashed red line), $G=\gamma$ (dotted-dashed blue line). }
\label{fig4}
\end{figure}

\begin{figure}[ht]
\includegraphics[width=0.5\columnwidth]{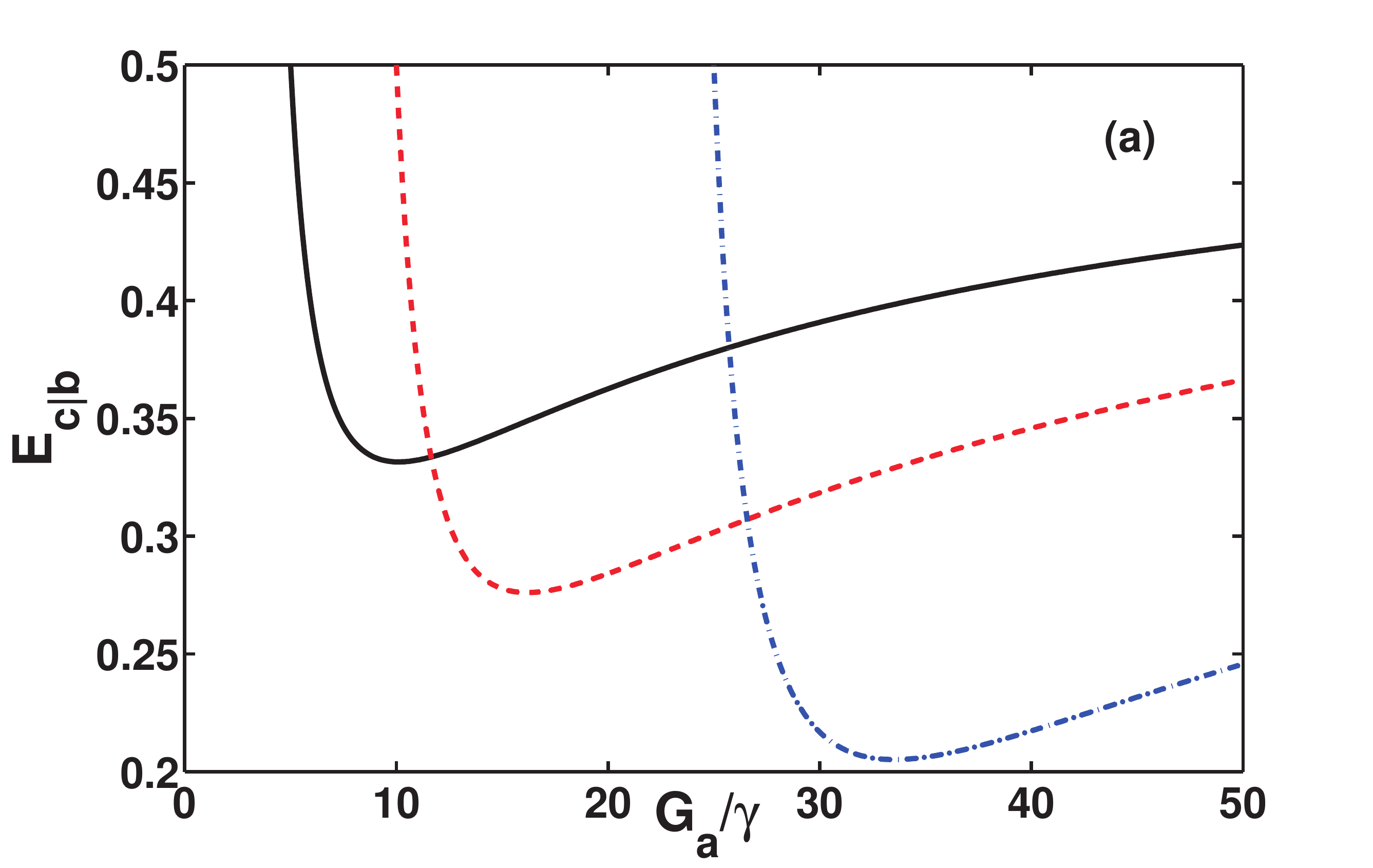}
\includegraphics[width=0.5\columnwidth]{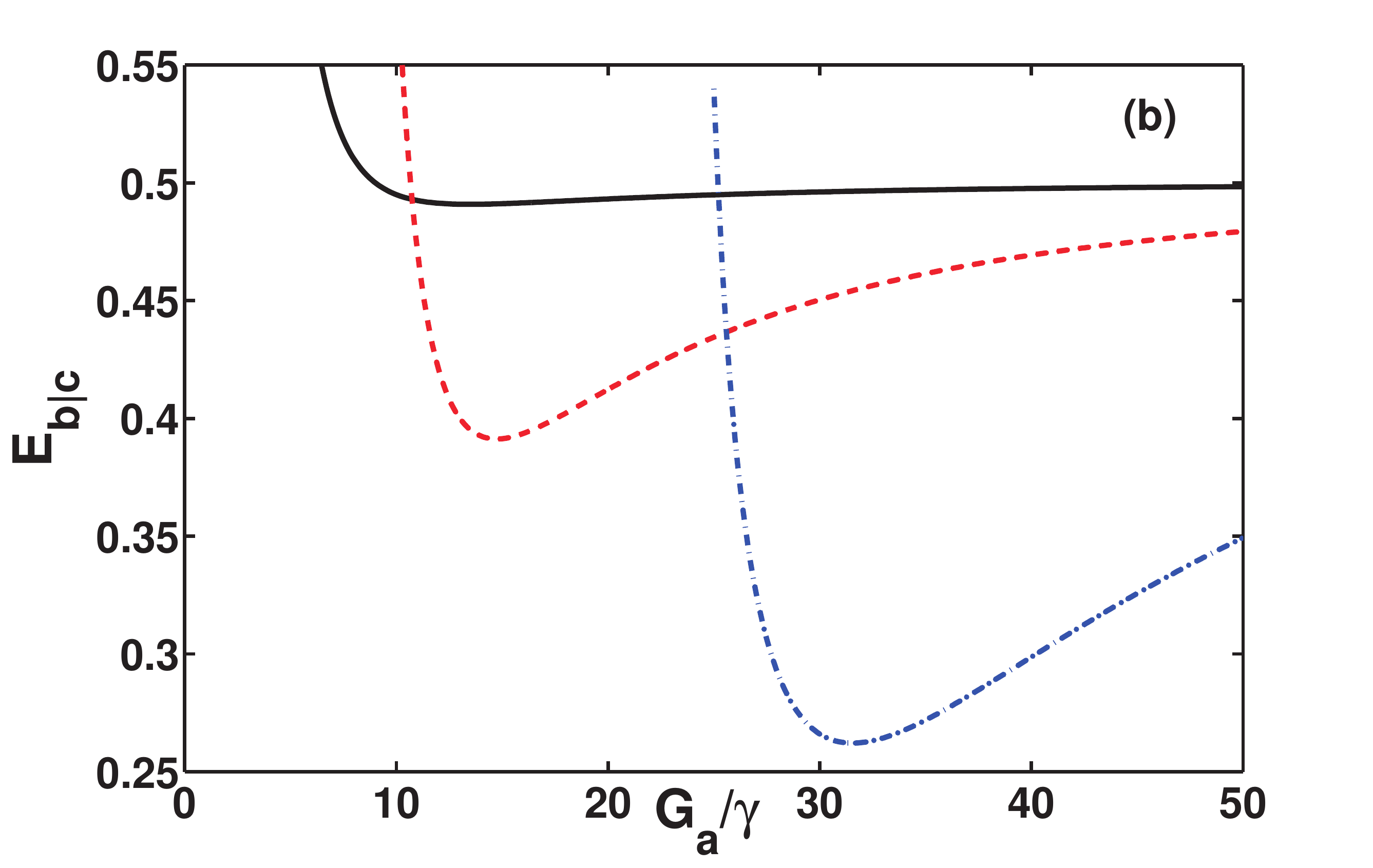}
\caption{The variation of the steering parameters (a) $E_{c|b}$ and (b) $E_{b|c}$ (right frame) with $G_{a}/\gamma$ for $n=n_{0}=0$ and several different values of the coupling constant $G>\gamma$: $G= 5\gamma$ (solid black line), $G=10\gamma$ (dashed red line), $G=25\gamma$ (dotted-dashed blue line). }
\label{fig5}
\end{figure}

The above analysis are illustrated in Figures~\ref{fig4} and~\ref{fig5}, where the steering parameter $E_{c|b}$ is shown in frames $(a)$, while $E_{b|c}$ is shown in frames (b) as a function of $G_{a}$ in the absence of the thermal excitations at the modes, $n=n_{0}=0$. The parameter regimes used in the plots were carefully chosen to satisfy the conditions for the stabile steady-state solutions.
Figure~\ref{fig4} shows the steering parameters for weak couplings $G\leq \gamma$, whereas Figure~\ref{fig5} shows the steering parameters for strong couplings $G>\gamma$. We see that in the case of $G\leq \gamma$ only one-way steering is observed that $E_{c|b}$ can be reduced below $1/2$ with $E_{b|c}$ always greater than $1/2$. The steering parameters for $G>\gamma$ are shown in Figure~\ref{fig5}. It is clear that we have a very different situation compared to that shown in Figure~\ref{fig4}. Now both $E_{c|b}$ and $E_{b|c}$ can be reduced below $1/2$ indicating that two-way steering becomes possible, meaning that the atomic and mirror modes can steer each other.
Thus, the creation of two-way steering requires larger values of $G$, that a strong nonlinear coupling of the cavity mode to the mirror mode is required to observe two-way steering. Note a significant asymmetry between $E_{c|b}$ and $E_{b|c}$. The asymmetry can be interpreted as arising from the asymmetry of the variances of the quadrature components of the modes. It is easily seen from Equation~(\ref{l10}) that $\Delta^{2}X_{b}\neq \Delta^{2}X_{c}$. Thus, we may conclude that the asymmetry of the steering is due to the asymmetry of the fluctuations the two modes.

\begin{figure}[ht]
\includegraphics[width=0.5\columnwidth]{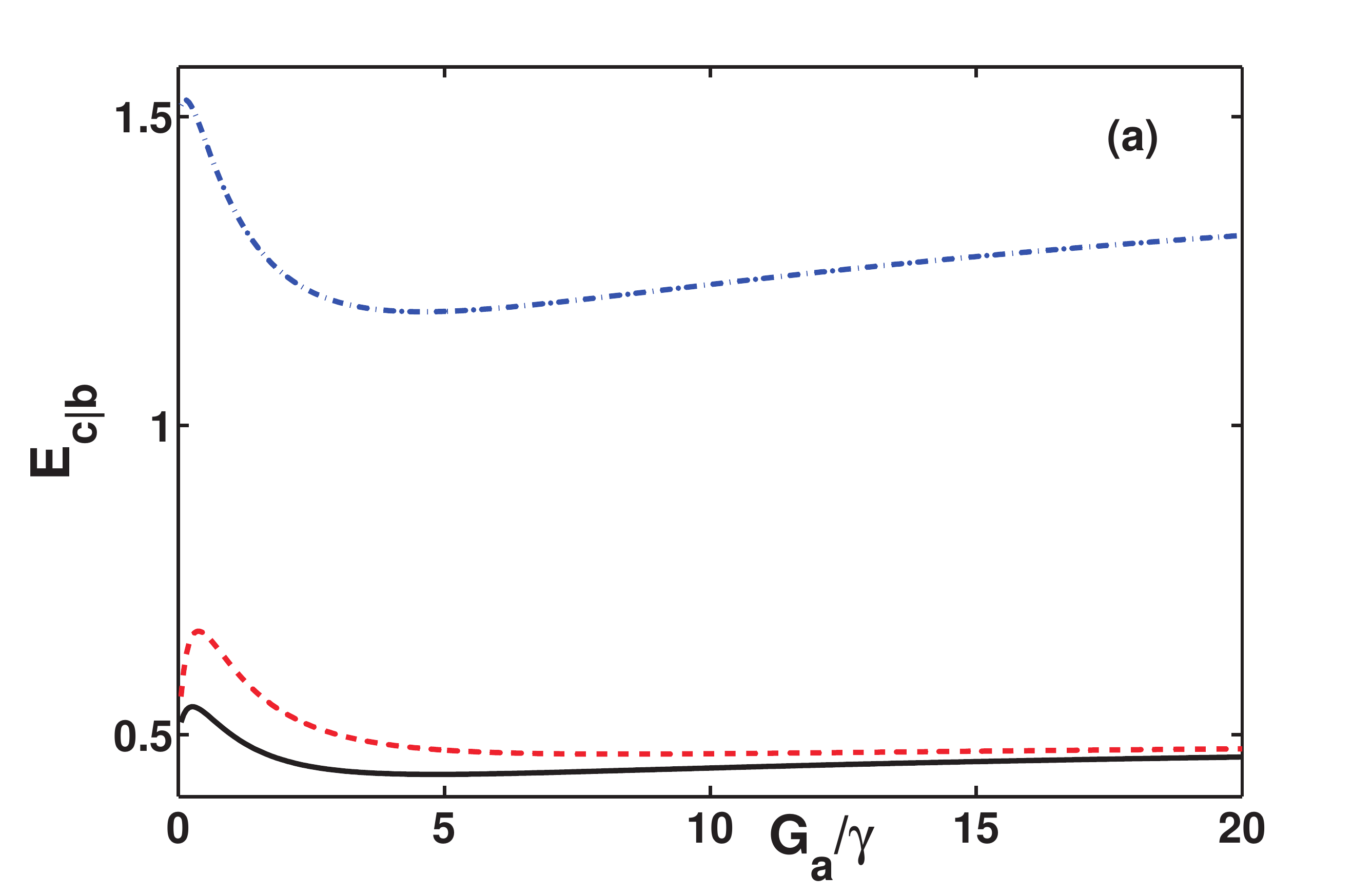}
\includegraphics[width=0.5\columnwidth]{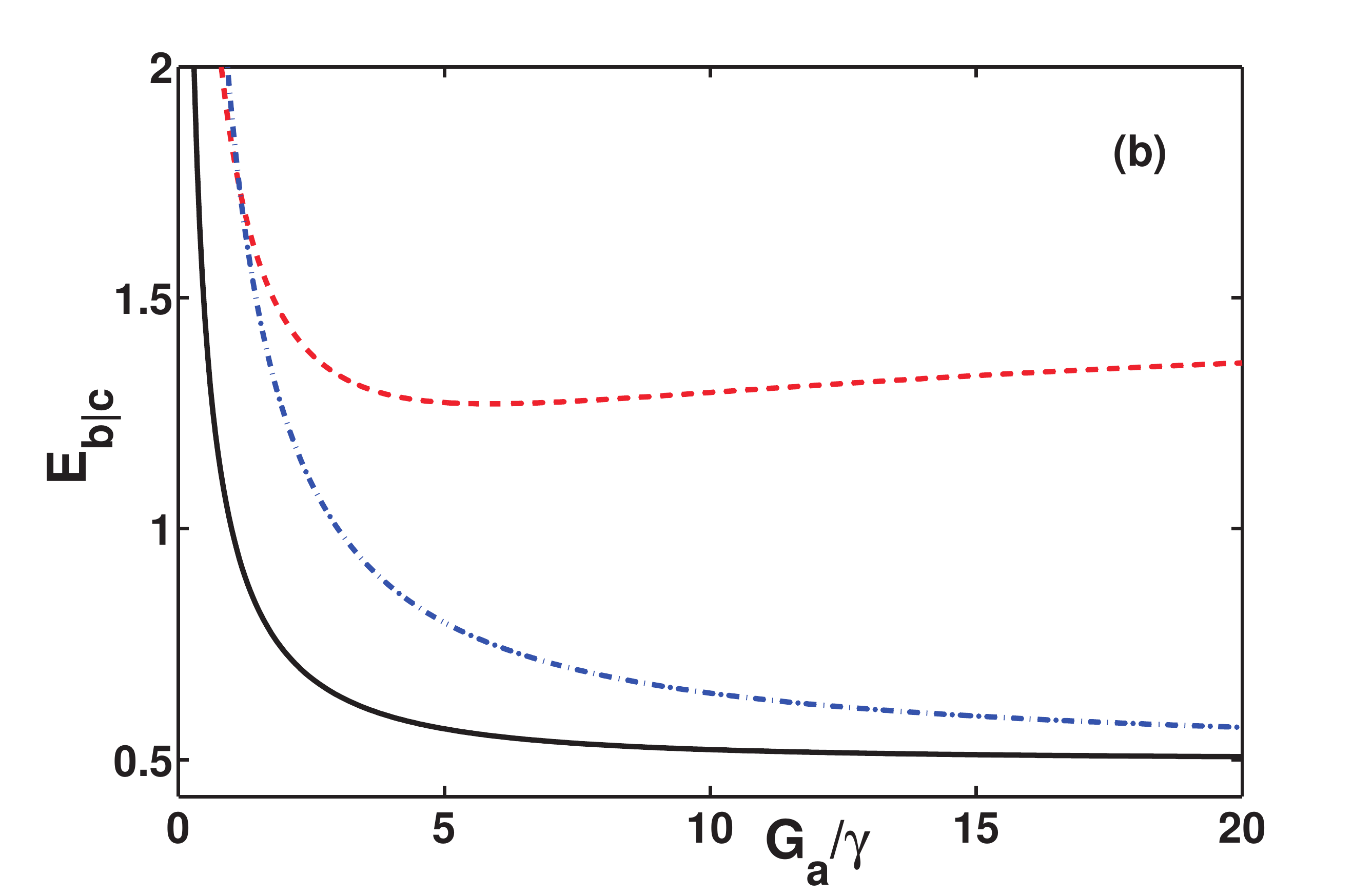}
\caption{Illustration of the effect of an asymmetric thermal excitation on the variation of the steering parameters (a) $E_{c|b}$ and (b) $E_{b|c}$ with $G_{a}/\gamma$ for $G=\gamma$. Here, the solid black line is for $(n,n_{0})=(0,0)$, dashed red line is for $(n,n_{0})=(0,1)$, and dotted-dashed blue line is for $(n,n_{0})=(1,0)$. For $(n,n_{0}) =(0,0)$ and $(0,1)$, $\Delta^{2}X_{b}>\Delta^{2}X_{c}$ while for $(n,n_{0}) =(1,0)$, $\Delta^{2}X_{b}<\Delta^{2}X_{c}$. }
\label{fig6}
\end{figure}

Figure~\ref{fig6} shows the sensitivity of the steering parameters to the thermal excitation of the modes. We consider asymmetric thermal excitations for which we get either $\Delta^{2}X_{b}>\Delta^{2}X_{c}$ or $\Delta^{2}X_{b}< \Delta^{2}X_{c}$. It is easily checked that for the thermal excitations with $(n,n_{0})=(0,0)$ and $(0,1)$, we have $\Delta^{2}X_{b}>\Delta^{2}X_{c}$, whereas with $(n,n_{0})=(1,0)$, we have $\Delta^{2}X_{b}< \Delta^{2}X_{c}$. It is seen from Figure~\ref{fig6}(a) that one-way quantum steering $(E_{c,b}<1/2)$ occurs even in the presence of the thermal excitation at which $\Delta^{2}X_{b}> \Delta^{2}X_{c}$. With the thermal excitation $(n,n_{0})=(1,0)$ at which the variance of the steered mode $c$ is larger than the steering mode $b$, $\Delta^{2}X_{c}> \Delta^{2}X_{b}$, the steering is destroyed completely. Similarly, one can see from frames (b) that the steering parameter $E_{b|c}$ is significantly less affected by the thermal excitation when $\Delta^{2}X_{c}> \Delta^{2}X_{b}$, i.e., when the variance of the steering mode $c$ is larger than the variance of the steered mode $b$. Clearly, the asymmetry of the steering is due to the asymmetry of the variances of the quadrature components of the two modes.
One can also see from Figure~\ref{fig6} that at large coupling strengths, $G_{a}\gg\gamma$, the steering parameters become insensitive to the thermal excitation at the steering mode. The results of Figure~\ref{fig6} clearly show that adding thermal excitation to the steered mode significantly affects the quantum steering while adding the thermal excitation to the steering mode is not so dramatic. In fact, the sensitivity of the steering parameters to the thermal excitation is not so dramatic if the variance of the steering mode is larger that that of the steered mode.

\section{Effect of the auxiliary mode on bipartite entanglement}\label{Sec5}

It should be pointed out that the role of the auxiliary mode in the creation of quantum steering contrasts with that for the creation of entanglement. To illustrate this point better, we now examine the asymmetric criterion for entanglement (\ref{l20}). The modes are entangled if the condition $\Delta_{a,b}^{h}<1$ is satisfied. Since in the case considered here $\Delta^{2} P_{a}=\Delta^{2} X_{a}$, $\Delta^{2} P_{b}=\Delta^{2} X_{b}$ and $\langle X_{a}P_{b}\rangle=\langle X_{b}P_{a}\rangle$, the expression (\ref{l20}) simplifies to
\begin{eqnarray}
\Delta_{a,b}^{h} = \frac{2\Delta^{2}\!\left(X_{a}+hP_{b}\right)}{1+h^{2}} .\label{l21}
\end{eqnarray}

In order to find the value of the parameter $h$ which minimizes $\Delta_{a,b}^{h}$, we take the derivative of $\Delta_{a,b}^{h}$ over $h$ and setting $\partial\Delta_{a,b}^{h}/\partial h=0$, we then arrive to a quadratic equation
\begin{equation}
\langle X_{a}P_{b}\rangle +\left(\Delta^{2} P_{b} -\Delta^{2} X_{a}\right)h -\langle X_{a}P_{b}\rangle h^{2} = 0 ,\label{l22}
\end{equation}
from which we find that the value of $h$ which minimizes $\Delta_{a,m}^{h}$ is
\begin{equation}
h = \frac{\left(\Delta^{2} P_{b} -\Delta^{2} X_{a}\right)- \sqrt{\left(\Delta^{2} P_{b} -\Delta^{2} X_{a}\right)^{2} +4\langle X_{a}P_{b}\rangle^{2}}}{2\langle X_{a}P_{b}\rangle} .\label{l23}
\end{equation}

\begin{figure}[ht]
\begin{centering}
\includegraphics[width=0.5\columnwidth]{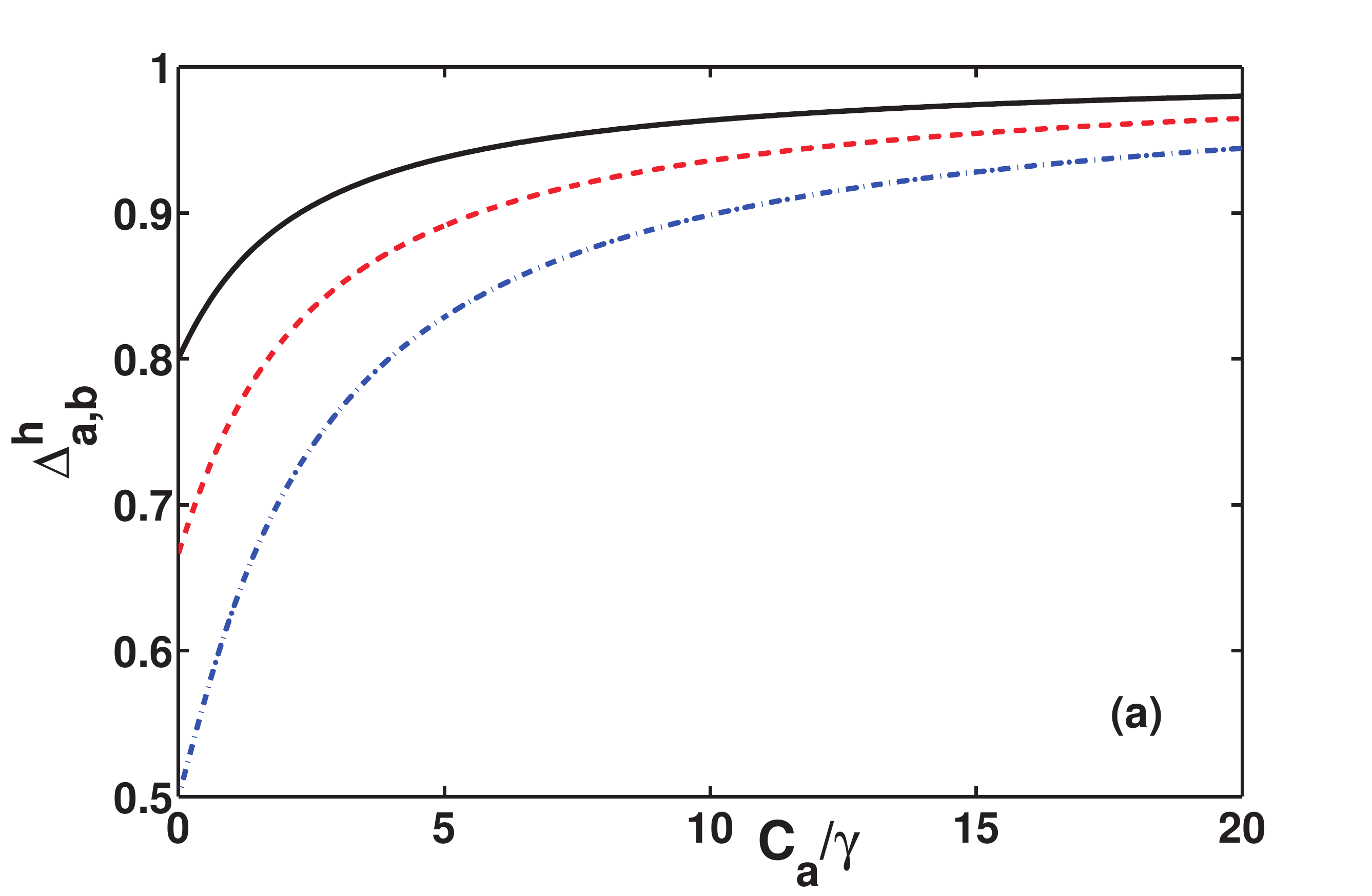}
\includegraphics[width=0.5\columnwidth]{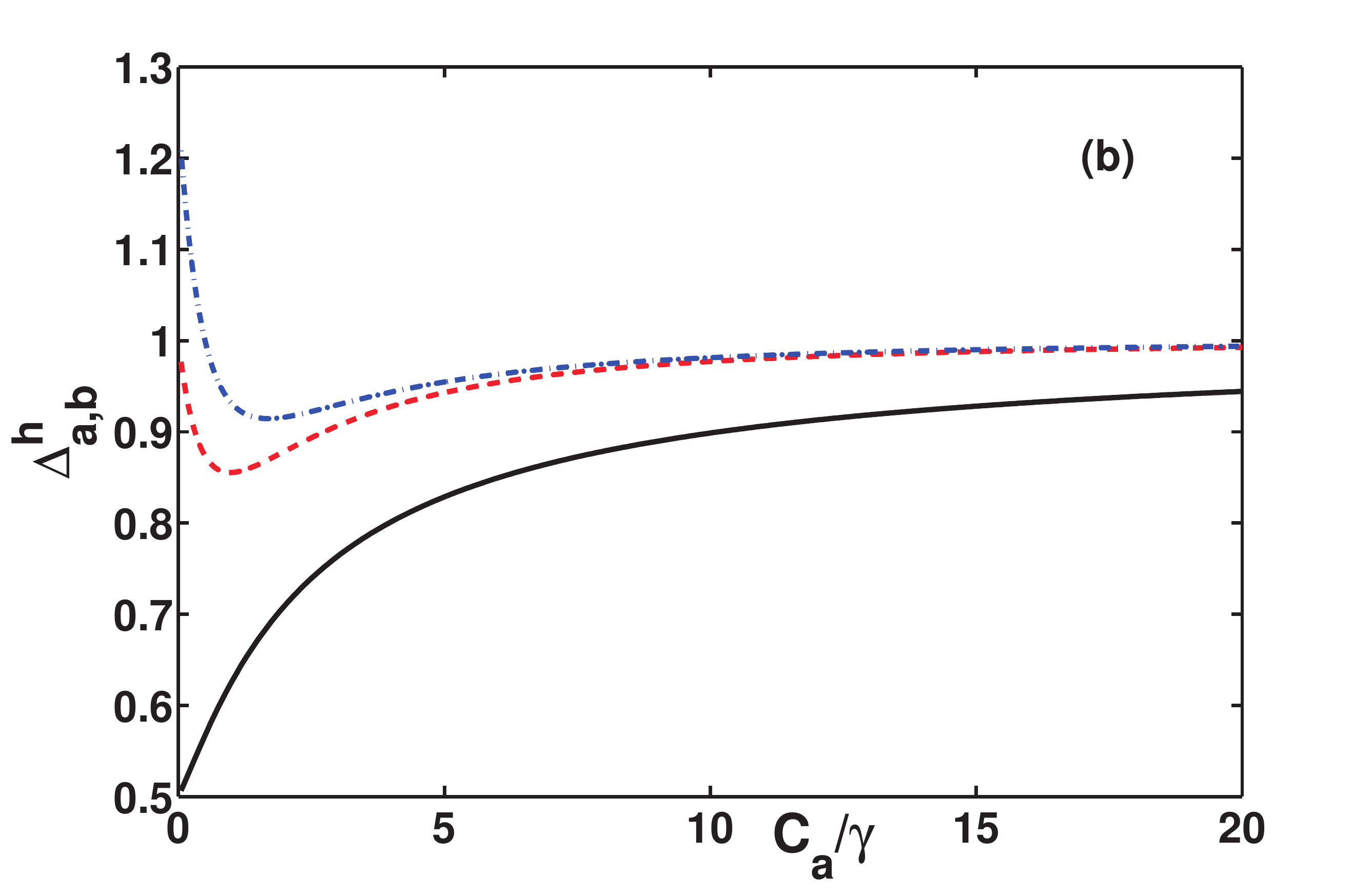}
\end{centering}
\caption{(a) Illustration of the effect of the coupling constant $g_{m}$ on the variation of the entanglement parameter $\Delta_{a,b}^{h}$ with $C_{a}$ in the case no thermal excitation at the modes $(n=n_{0}=0)$. Solid black line corresponds to $g_{m}=0.25\gamma$, dashed red line to $g_{m}=0.5\gamma$, and dotted-dashed blue line to $g_{m}=\gamma$. (b) Illustration of the effect of the thermal excitation at the mirror mode on the variation of $\Delta_{a,b}^{h}$ with $C_{a}$ in the case no thermal excitation at the cavity mode $(n=0)$ and $g_{m}=\gamma$. Solid black line corresponds to $n_{0}=0$, dashed red line to $n_{0}=1$, and dotted-dashed blue line to $n_{0}=1.5$. }
\label{fig3}
\end{figure}

Figure~\ref{fig3} shows the variation of the entanglement parameter $\Delta_{a,b}^{h}$, as given in Equation~(\ref{l21}), with the coupling strength $C_{a}$ for various coupling strengths $g_{m}$ between the modes and various thermal excitations at the mirror mode. We see from Figure~\ref{fig3}(a) that where there is no thermal excitation, the modes are entangled in the absence of the fast damped auxiliary mode and the entanglement decays steadily with $C_{a}$ but is never destroyed completely.
The effect of the thermal excitation at the mirror mode shown in Figure~\ref{fig3}(b) is to reduce the entanglement at small $C_{a}$. It is interesting to note that for $n_{0}<1$ the entanglement is present over the entire range of $C_{a}$. It starts to be destroyed for $n_{0}>1$. Moreover, in the absence of the auxiliary mode $(C_{a}=0)$ the entanglement is much faster destroyed by the thermal excitation than in the presence of the auxiliary mode.

The results of Figures~\ref{fig2} and \ref{fig3} clearly show that in the absence of the auxiliary mode the parametric-type interaction between two equally damped modes $(\gamma_{m}=\kappa\equiv\gamma)$ can generate steady-state entanglement but cannot generate quantum steering between the modes. We may conclude that it is possible to entangle two modes  even if the variances of the quadrature components of the modes are equal, but the generation of quantum steering requires the variances to be unequal.
Moreover, the results show that the fast damped auxiliary mode has a constructive effect on quantum steering of the parametrically coupled modes despite its destructive effect on entanglement between these modes, seen in Figure~\ref{fig3}.

\section{Summary}\label{Sec6}

In summary, we have shown that a fast damped mode of a three-mode system can create quantum steering between the remaining two modes. The discussion has concentrated on a three-mode optomechanical system composed of a two-level atom located inside a single mode cavity with a movable mirror. We have considered two cases. In the first, we have assumed that the atomic mode, which directly couples to the cavity mode, appears as a fast damped "auxiliary" mode. In the second, we have taken the cavity mode as a fast damped "auxiliary" mode which directly couples to both atomic and mirror modes.
We have presented analytical solutions for the steering parameters of the Reid criterion and have found that one-way steering always occurs when an auxiliary mode is coupled to one of the modes of the system, and two-way steering requires the simultaneous coupling of the auxiliary mode to both modes.
We have found that that the cavity and mirror modes, which are coupled by the parametric-type interaction, can be entangled but cannot exhibit quantum steering. When, in addition, the cavity mode is coupled to a fast damped atomic mode, steering correlations are created and the modes then exhibit one-way quantum steering. When the cavity mode appears as a fast damped mode, the atomic and mirror mode may exhibit two-way steering. We have demonstrated that the auxiliary mode creates an asymmetry between the variances of the quadrature components of the modes. We have found that the asymmetry is crucial for the generation of the steering correlations and quantum steering may occur only when the variance of the steering mode is larger that the variance of the steered mode. We have also discussed the effect of the thermal excitation of the modes and show that the scheme is quite robust against the thermal excitation of the modes if the fluctuations of the steering mode are larger than the fluctuations of the steered mode.

\ack
This work was supported by National Science Foundation (NSF) (11304024, 11547007, 61505014); Yangtze Fund for Youth Teams of Science and Technology Innovation (2015cqt03).


\section*{References}

\begin{thebibliography}{70}


\bibitem{es35} Sch\"{o}dinger E 1935  {\it Proc. Cambridge Philos. Soc.} {\bf 31} 555

\bibitem{wj07} Wiseman H M, Jones S J and Doherty A C 2007 {\it Phys. Rev. Lett.} \textbf{98} 140402

\bibitem{jw07} Jones S J, Wiseman H M and Doherty A C 2007 {\it Phys. Rev.} A \textbf{76} 052116

\bibitem{cj09} Cavalcanti E G, Jones S J, Wiseman H M and Reid M D 2009 {\it Phys. Rev.} A \textbf{80} 032112

\bibitem{rd09} Reid M D, Drummond P D, Bowen W P, Cavalcanti E G, Lam P K, Bachor H A,  Andersen U L and Leuchs G 2009 {\it Rev. Mod. Phys.} \textbf{81} 1727

\bibitem{rt15} Rosales-Zarate L, Teh R Y, Kiesewetter S, Brolis A, Ng K and Reid M D 2015 {\it J. Opt. Soc. Am.} B \textbf{32} A82
    
\bibitem{gsa14} Chowdhury P, Pramanik T, Majumdar A S and Agarwal G S 2014 {\it Phys. Rev.} A \textbf{89} 012104

\bibitem{bc} Branciard C, Cavalcanti E G, Walborn S P, Scarani V and Wiseman H M 2012  {\it Phys. Rev.} A \textbf{85} 010301(R)

\bibitem{md13} Reid M D 2013 {\it Phys. Rev.} A {\bf 88} 062338

\bibitem{ww14} Walk N, Hosseni S, Geng J, Thearle O, Haw J Y, Armstrong S, Assad S M, Janousek J, Ralph T C, Symul T, Wiseman H M and Lam P K 2016 {\it Optica} {\bf 3} 634

\bibitem{sd13} Schneeloch J, Dixon P B, Howland G A, Broadbent C J and Howell J C 2013 {\it Phys. Rev. Lett.} \textbf{110} 130407

\bibitem{olsen13} Olsen M K and Corney J F 2013 {\it Phys. Rev.} A \textbf{87} 033839

\bibitem{sn14} Skrzypczyk P, Navascu\'{e}s M and Cavalcanti D 2014 {\it Phys. Rev. Lett.} \textbf{112} 180404

\bibitem{bv14} Bowles J, V\'{e}rtesi T, Quintino M T and Brunner N 2014 {\it Phys. Rev. Lett.} \textbf{112} 200402

\bibitem{tr14} Jevtic S, Pusey M, Jennings D and Rudolph T 2014 {\it Phys. Rev. Lett.} \textbf{113} 020402

\bibitem{wgh14} Wang M, Gong Q and He Q Y 2014 {\it Opt. Lett.} \textbf{39} 6703

\bibitem{he15} He Q Y, Gong Q H and Reid M D 2015 {\it Phys. Rev. Lett.} {\bf 114} 060402

\bibitem{ade14} Kogias I, Lee A R, Ragy S and Adesso G 2015 {\it Phys. Rev. Lett.} {\bf 114} 060403

\bibitem{ade15} Kogias I and Adesso G 2015 {\it J. Opt. Soc. Am.} B {\bf 32} A27

\bibitem{rh16}  Rao S, Hu X, Li L C and Xu J 2016 {\it Optics Express} {\bf 25} 11584

\bibitem{zc17} Zhong W X, Cheng G L and Hu X M 2017 {\it Optics Express} {\bf 25} 11584

\bibitem{wl17} Wang L, Lv S and Jing J 2017  {\it Opt. Express} {\bf 25} 17457

\bibitem{kl17} Kalaga J K, Leo\'nski W and Szczesniak R 2017 {\it Quant. Inf. Proc.} {\bf 16} 265

\bibitem{kl17a} Kalaga J K, Leo\'nski W and Szczesniak R 2017 {\it Photonics Lett.} {\bf 9} 97

\bibitem{ct18} Cheng G L, Tan H and Chen A 2018 {\it Optics Commun.} {\bf 412} 166

\bibitem{wx18} Wang F, Xu J, Cheng G L and Oh C H 2018 {\it Ann. Phys.} {\bf 388} 162

\bibitem{sj10} Saunders D J, Jones S J, Wiseman H M and Pryde G J 2010 {\it Nature Phys.} \textbf{6} 845

\bibitem{sg12} Smith D H, Gillett G, de Almeida M P, Branciard C, Fedrizzi A, Weinhold T J, Lita A, Calkins B, Gerrits T, Wiseman H M, Nam S W and White A G 2012 {\it Nature Commun.} \textbf{3} 625

\bibitem{be12} Bennet A J, Evans D A, Saunders D J, Branciard C, Cavalcanti E G, Wiseman H M and Pryde G J 2012 {\it Phys. Rev.} X \textbf{2} 031003

\bibitem{vh12}  H\"{a}ndchen V, Eberle T, Steinlechner S, Samblowski A, Franz T, Werner R F and Schnabel R 2012 {\it Nature Photonics} \textbf{6} 596

\bibitem{bw12} Wittmann B, Ramelow S, Steinlechner F, Langford N K, Brunner N, Wiseman H M, Ursin R and Zeilinger A 2012 {\it New J Phys.} \textbf{14} 053030

\bibitem{ss13} Steinlechner S, Bauchrowitz J, Eberle T and Schnabel R 2013 {\it Phys. Rev.} A \textbf{87} 022104

\bibitem{Guo14} Sun K, Xu J S, Ye X J, Wu Y C, Chen J L, Li C F and Guo G C 2014 {\it Phys. Rev. Lett.} \textbf{113} 140402

\bibitem{nc15} Kocsis S, Hall M J W, Bennet A J, Saunders D J and Pryde G J 2015 {\it Nature Commun.} \textbf{6} 5886

\bibitem{seiji15} Armstrong S, Wang M, Teh R Y, Gong Q H, He Q Y, Janousek J, Bachor H A, Reid M D and Lam P K 2015 {\it Nature Phys.} {\bf11} 167

\bibitem{ww16} Wollmann S, Walk N, Bennet A J, Wiseman H M and Pryde G J 2016 {\it Phys. Rev. Lett.} {\bf 116} 160403

\bibitem{wc14} Woolley M J and Clerk A A 2014 {\it Phys. Rev.} A {\bf 89} 063805

\bibitem{tz15} Tan H, Zhang X and Li G X 2015 {\it Phys. Rev.} A {\bf 91} 032121

\bibitem{yl15} Yan Y, Li G X and Wu Q L 2015 {\it Opt. Express} {\bf 23} 21306

\bibitem{xs15} Xiang Y, Sun F X, Wang M, Gong Q H and He Q Y 2015 {\it Opt. Express} {\bf 23} 30104

\bibitem{ts15} Tan H and Sun L H 2015 {\it Phys. Rev.} A  {\bf 92} 063812

\bibitem{qd17} El Qars J, Daoud M and Laamara R A 2017 {\it Eur. Phys. J.} D {\bf 71} 122

\bibitem{lz17} Li J and Zhu S Y 2017 {\it Phys. Rev.} A {\bf 96} 062115

\bibitem{vt07} Vitali D, Tombesi P, Woolley M J, Doherty A C and Milburn G J 2007 {\it Phys. Rev.} A {\bf 76} 042336

\bibitem{gv07} Genes C, Vitali D and Tombesi P 2008 {\it Phys. Rev.} A {\bf 77} 050307(R)

\bibitem{vg07} Vitali D, Gigan S, Ferreira A, B\"{o}hm H R, Tombesi P, Guerreiro A, Vedral V, Zeilinger A and Aspelmeyer M 2007 {\it Phys. Rev. Lett.} {\bf 98} 030405

\bibitem{bg08} Bhattacharya M, Giscard P L and Meystre P 2008 {\it Phys. Rev.} A {\bf 77} 013827

\bibitem{gm08} Genes C, Mari A, Tombesi P and Vitali D 2008 {\it Phys. Rev.} A {\bf 78} 032316

\bibitem{zh11} Zhou L, Han Y, Jing J and Zhang W 2011 {\it Phys. Rev.} A {\bf 83} 052117

\bibitem{sl12} Sun L H, Li G X and Ficek Z 2012 {\it Phys. Rev.} A \textbf{85} 022327

\bibitem{wc13} Wang Y D and  Clerk A A 2013 {\it Phys. Rev. Lett.} \textbf{110} 253601

\bibitem{sx17} Sun L H, Xia F, Zhang K K, Zhang H F and Xu D H 2017 {\it Eur. Phys. J.} D {\bf 71} 311

\bibitem{hw11} Hofer S G, Wieczorek W, Aspelmeyer M and Hammerer K 2011 {\it Phys. Rev.} A \textbf{84} 052327

\bibitem{hr13} He Q Y and Reid M D 2013 {\it Phys. Rev.} A \textbf{88} 052121

\bibitem{hf14} He Q Y and Ficek Z 2014 {\it Phys. Rev.} A \textbf{89} 022332

\bibitem{wg14} Wang M, Gong Q, Ficek Z and He Q Y 2014 {\it Phys. Rev.} A \textbf{90} 023801

\bibitem{kh14} Kiesewetter S, He Q Y, Drummond P D and Reid M D 2014 {\it Phys. Rev.} A {\bf 90} 043805

\bibitem{wg14a} Wang M, Gong Q, Ficek Z and He Q Y 2015 {\it Sci. Rep.} \textbf{5} 12346

\bibitem{sm17} Sun F X, Mao D, Dai Y T, Ficek Z, He Q Y and Gong Q H 2017 {\it New J. Phys.} {\bf 19} 123039

\bibitem{md89} Reid M D 1989 {\it Phys. Rev.} A {\bf 40} 913

\bibitem{ps06} Parkins A S, Solano E and Cirac J I 2006 {\it Phys. Rev. Lett.} {\bf 96} 053602

\bibitem{ph99} Plenio M B, Huelga S F, Beige A and Knight P L 1999 {\it Phys. Rev.} A {\bf 59} 2468

\bibitem{gb06} Gigan S, Bohm H R, Paternostro M, Blaser F, Langer G, Hertzberg J B, Schwab K, Baeuerle D, Aspelmeyer M and Zeilinger A 2006 {\it Nature} {\bf 444} 67

\bibitem{gh09} Groblacher S, Hammerer K, Vanner M R and Aspelmeyer M 2009 {\it Nature} {\bf 460} 724

\bibitem{rd11} Riviere R, Deleglise S, Weis S, Gavartin E, Arcizet O, Schliesser A and Kippenberg T J 2011 {\it Phys. Rev.} A {\bf 83} 063835

\bibitem{c11} Chan J, Mayer Alegre T P, Safavi-Naeini A H, Hill J T, Krause A, Groblacher S, Aspelmeyer M and Painter O 2011 {\it Nature }{\bf 478} 89

\bibitem{td11} Teufel J D, Donner T, Li D, Harlow J H, Allman M S, Cicak K, Sirois A J, Whittaker J D, Lehnert K W and Simmonds R W 2011 {\it Nature }{\bf 475} 359

\bibitem{rn10} Rocheleau T, Ndukum T, Macklin C, Hertzberg J B, Clerk A A and Schwab K C 2010 {\it Nature} {\bf 463} 72

\bibitem{hp40} Holstein T and Primakoff H 1940 {\it Phys. Rev.} {\bf 58} 1098

\bibitem{gm01} Giovannetti V, Mancini S and Tombesi P 2001 {\it Europhys. Lett.} {\bf 54} 559

\bibitem{mg02}  Mancini S, Giovannetti V, Vitali D and Tombesi P 2002 {\it Phys. Rev. Lett.} {\bf 88} 120401

\bibitem{pp13} Purdy T P, Peterson R W and Regal C A 2013 {\it Science} {\bf 339} 801

\bibitem{pp16} Peterson R W, Purdy T P, Kampel N S, Andrews R W, Yu P L, Lehnert K W and Regal C A 2016 {\it Phys. Rev. Lett.} {\bf 116} 063601

\bibitem{dg00} Duan L M, Giedke G, Cirac J I and Zoller P 2000 {\it Phys. Rev. Lett.} {\bf 84} 2722

\bibitem{rs00} Simon R 2000 {\it Phys. Rev. Lett.} {\bf 84} 2726

\end{thebibliography}
\end{document}